\documentclass[journal]{IEEEtran}
\newcommand{\field}[1]{\mathbb{#1}} 

\newcommand{\trace}[1]{\textrm{\textbf{tr}} \left ( {#1} \right )}

\newcommand{\diag}[1]{\textrm{\textbf{ diag}} \left ( {#1} \right )}

\newcommand{\BIGO}[1]{\textrm{O} \left ( {#1} \right )}

\newcommand{\pderiv}[2]{\frac{\partial {#1}}{\partial {#2}} }

\newcommand{\ALGCMT}[1]{$\quad \backslash \backslash$ {#1}}
\newcommand{\BRAK}[1]{\left [ {#1} \right ]}
\newcommand{\CBRAK}[1]{\left \{ {#1} \right \} }
\newcommand{\ABS}[1]{\left | {#1} \right | }
\newcommand{\PAREN}[1]{\left ( {#1} \right )}

\newcommand{\dist}[1]{\textrm{dist} \left ( {#1} \right )}

\newcommand{\LTWOF}[3]{\textbf{$L^2_{#3}$}\PAREN{{#1}, {#2}}}

\newcommand{\NORM}[1]{\left \| {#1} \right \| }
\newcommand{\EX}[1]{\mathbb{E} \left [ {#1} \right ] }

\newcommand{\FLOOR}[1]{\left \lfloor {#1} \right \rfloor}

\ifCLASSINFOpdf
\else
\fi

\usepackage[cmex10]{amsmath}
\usepackage{hyperref} 
\usepackage{graphicx, epsfig}
\usepackage{amsfonts}
\usepackage{algorithm, algorithmic}
\usepackage{bm}
\usepackage{mathptmx}

\usepackage[font={small}]{caption, subfig}

 \renewcommand{\figurename}{Fig.}
 \renewcommand{\tablename}{Tab.}

\begin{document}
%
\title{Gaussian Process Models for HRTF based Sound-Source Localization and Active-Learning}
%
%
%

\author{
        Yuancheng Luo, ~\IEEEmembership{Member,~IEEE}, Dmitry N. Zotkin, Ramani Duraiswami, ~\IEEEmembership{Member,~IEEE}
        \thanks{Yuancheng Luo, Dmitry N. Zotkin, and Ramani Duraiswami are with the University of Maryland Center for Advanced Computer Studies in College Park, 20742 USA, e-mail: yluo1@umd.edu, dz@umiacs.umd.edu, ramani@umiacs.umd.edu.}\\ 
}

\markboth{IEEE Journal of Selected Topics in Signal Processing, March~2014}%
{Shell \MakeLowercase{\textit{et al.}}: Bare Demo of IEEEtran.cls for Journals}
%



\maketitle

\begin{abstract}

From a machine learning perspective, the human ability localize sounds can be modeled as a non-parametric and non-linear regression problem between binaural spectral features of sound received at the ears (input) and their sound-source directions (output). The input features can be summarized in terms of the individual's head-related transfer functions (HRTFs) which measure the spectral response between the listener's eardrum and an external point in $3$D.  Based on these viewpoints, two related problems are considered: how can one achieve an optimal sampling of measurements for training sound-source localization (SSL) models, and how can SSL models be used to infer the subject's HRTFs in listening tests. First, we develop a class of binaural SSL models based on Gaussian process regression and solve a \emph{forward selection} problem that finds a subset of input-output samples that best generalize to all SSL directions. Second, we use an \emph{active-learning} approach that updates an online SSL model for inferring the subject's SSL errors via headphones and a graphical user interface. Experiments show that only a small fraction of HRTFs are required for $5^{\circ}$ localization accuracy and that the learned HRTFs are localized closer to their intended directions than non-individualized HRTFs.

\end{abstract}


%
\IEEEpeerreviewmaketitle

\section{Introduction}
\label{SEC:GPAL:INTRO}

Many animals possess a remarkable omnidirectional sound localization ability enabled by subconsciously processing subtle features in the sounds received at the two ears from a common source location. For humans, these features arise due to the incoming acoustic wave scattering off the listener's anatomical features (head, torso, pinnae) before reaching the eardrum. The spectral ratio between the sounds recorded at the eardrum and that would have been obtained at the center of the head in absence of the listener is known as the head-related transfer function (HRTF) \cite{BLAUERT}; HRTFs are thus specific to the individual's anthropometry, wave direction, and contain other important cues such as the interaural time delay (ITD) and the interaural level difference (ILD) \cite{CHENG1999}. Moreover, knowledge of individualized HRTFs allow for perceptually accurate $3$D spatial audio synthesis \cite{KULKARNI1998,WENZEL2,ROMIGH}. 

We investigate the \emph{pre-image} problem, namely how pairs of left and right ear HRTFs and functions of HRTFs (features based on them) map back to their measurement directions. This is related to the problem of sound-source localization (SSL) where under simple (anechoic) conditions, the direction of an acoustic event can be inferred from multi-receiver recordings of the sound spectrum by expressing the spectral cues solely in terms of the receiver's transfer functions (independent of their actual content). This is of interest in robot perception (e.g. for event detection and localization \cite{HORNSTEIN,ROTHBUCHER}), where the receiver's transfer functions can be measured beforehand. For humans, this problem is restricted to two receivers (human ears) where functions of left and right pairs of HRTFs are mapped to their measurement directions in place of SSL directions. Thus, it possible to model this relation as either a classification or a regression problem between the two domains. Many works in literature have attempted similar tasks.

\subsection{Prior Works}
Cue-mapping \cite{RODEMANN} uses ITD, ILD, and interaural envelope difference features paired with azimuth directions in a weighted kernel nearest-neighbor (NN) setting. A linear mapping between ITD, ILD, and HRTF notch frequency features to spherical coordinates can be learned \cite{HORNSTEIN}. A self-organizing map between input ITD, spectral notches features and output horizontal and median plane coordinates can be trained \cite{NAKASHIMA}. Conditional probability maps derived from per-frequency ITD and ILD can be used to estimate direction via a maximum a posteriori estimator \cite{WILLERT}. A probabilistic affine regression model between interaural transfer functions and the direction is possible \cite{DELEFORGE}.

Most closely related to our work are the source-cancellation and match-filtering algorithms \cite{KEYROUZ1,KEYROUZ2,KEYROUZ3,POURMOHAMMAD}, where the binaural recordings ($S_L$ left, $S_R$ right ears) are represented as convolutions of a common sound-source signal $S$ and the appropriate filters; for recording done in an anechoic space, these filters are the same-direction HRTFs ($H_L$ left, $H_R$ right ears). The per-frequency domain representation is given by
\begin{equation}\label{EQ:GPAL:INTRO:FREQCONV}
\displaystyle
\begin{split}
S_L & = H_L \circ S, \quad S_R=H_R \circ S, 
\end{split}
\end{equation}
where $\circ$ is element-wise product. The source-signal $S$ is removed by computing the ratio between left and right channel recordings ($\frac{S_L}{S_R} = \frac{H_L}{H_R}$). These binaural features, which are reduced to ratios of HRTFs, can be compared to those pre-computed from the subject's collection of measured HRTFs; the measurement direction belonging to the maximally cross-correlated pair is reported as the sound-source direction. Such an approach can be interpreted as a nearest neighbor (NN) classifier where the binaural features and measurement directions are single class instances and labels respectively. 

\subsection{Present Work}

We propose a generalization of the match-filtering algorithm that addresses several deficiencies: While an NN classifier is accurate for a large number of training samples, it does not report out-of-sample spatial directions unless specified in a regression context. Linear regression methods via ordinary least squares (OLS) regressors\footnotemark\footnotetext{$\textbf{y} = \textbf{x}^T \beta, \quad \beta = \PAREN{X^T X}^{-1} X^T Y$, for parameters $\beta$} often perform poorly due to inaccurate assumptions on the model complexity (number of parameters) and the linearity between predictors and outputs. Common issues include over-fitting the model to noise that arise from parametric OLS methods and under-fitting the training data from assumptions of linearity. Instead, we adopt a non-linear and non-parametric\footnotemark\footnotetext{Number of parameters is proportional to the number of data samples conditioned upon for inference.} Gaussian process (GP) regression (GPR) \cite{RASMUSSEN} framework to address these issues.

GPR is a \emph{kernel method}\footnotemark\footnotetext{Predictor variables are implicitly mapped to a reproducing kernel Hilbert space whose inner products are taken to be evaluations of a valid Mercer kernel or covariance function.} that places weak assumptions on the joint probability distribution\footnotemark\footnotetext{Normal distribution defined by prior mean and covariance functions of predictor variables (binaural features).} of \emph{latent function realizations} that would model the output observations (spatial directions) in a Bayesian setting. Observations are drawn (realized) from a high-dimensional normal distribution that represents the joint probability density function of a collection of random variables indexed by their predictor variables. GPs have several attractive properties that are well-suited for SSL.

Based on the observation that HRTFs corresponding to different spatial directions covary smoothly with the considered binaural features (see sections \ref{SEC:GPAL:FEATS}), we show they can be modeled via simple stationary GP covariance functions (see section \ref{SEC:GPAL:GPR}). The GP Bayesian formulation allows for the choice of the covariance function, which governs the smoothness between realizations at nearby predictors, to be automatically selected by evaluating a data marginal-likelihood criterion (goodness-of-fit); covariance functions belong to a function class and are specified by their ``hyperparameters'' (parameters that describe distributions). This allows the covariance function hyperparameters to be learned without the need for cross-validation and provides insights as to the intrinsic dimensionality of the high-dimensional feature space that the binaural features are mapped to. Most importantly, uncertainties  in GP prediction are well-defined in terms of both prior and posterior distributions; the predicted variances at different inputs are tractable. Thus, GPR generalizes NN classifiers as it makes non-linear inferences to observations outside the training set. By the representer theorem, kernel methods such as support vector regression (SVR) \cite{SMOLA} and GPR make predictions expressible as linear combinations of non-linear covariance evaluations between the training features/observations and the test features. 

In general, GPs perform better (make accurate inferences) with more observations (data) than other non-linear regression methods that do not encode and select for prior data-assumptions. The trade-off is its high computational costs ($\BIGO{N^3}$ operations for $N$ number of observations) for both model-selection and inference; scaling GPs for for large datasets is an active field of research. Fortunately, the availability of high quality datasets, computational resources, and faster algorithmic formulations have allowed us to overcome these problems.  In previous works, we have used several properties of HRTF datasets to to perform fast GP based HRTF interpolation \cite{LUO3} and data-fusion \cite{LUO5}. The current work is a major extension of our recent work on binaural SSL \cite{LUO7}. For future references, we refer to GPs that predict SSL directions as \textbf{GP-SSL} models (see section \ref{SEC:GPAL:GPR} for a complete derivation). 

\section{Formulation of Problems}
\label{SEC:GPAL:INTRO:PROB}

\begin{figure}[ht]
  \centering
\includegraphics[width=.45\textwidth]{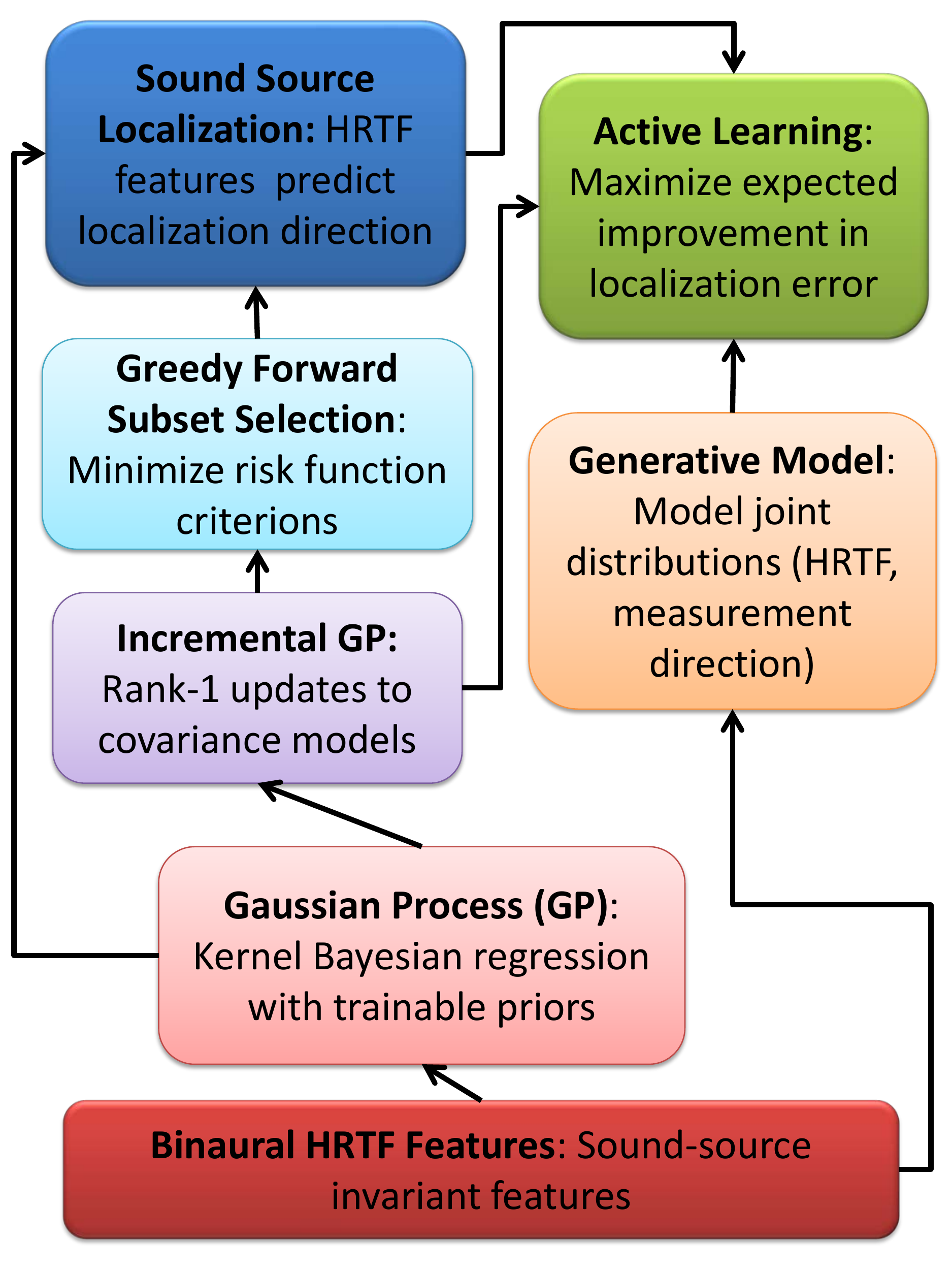}
\caption{Gaussian Process Regression with binaural features (bottom two boxes) to perform two types of inferences. On the left are shown the steps needed to perform sound-source localization. On the right is shown an active-learning framework that combines SSL with listening tests to learn a listener's HRTFs.
}
\label{FIG:GPAL:INTRO:CONCEPT}
\end{figure}

This work investigates two problems related to GP-SSL models (see Fig. \ref{FIG:GPAL:INTRO:CONCEPT} ). For notation, we refer to a binaural feature as a $D$-dimensional vector $\textbf{x} \in \field{R}^{D}$ ($D$ is number of frequency bins), the measurement direction as the unit vector $\textbf{y} \in \field{R}^{M}$ ($M=3$ for the standard Cartesian basis), and collections of the aforementioned quantities ($N$ number of samples) as concatenated into matrices $X \in \field{R}^{N \times D}$ and $Y \in \field{R}^{N \times M}$. The binaural features are independent of the sound-source content and thus strictly functions of the subject's HRTFs (see section \ref{SEC:GPAL:FEATS}). GP-SSL models are thereby specified and trained over known HRTFs and measurement directions belonging to CIPIC \cite{ALGAZI} database subjects. 

\subsection{Feature subset-selection}

Subset-selection for non-parametric methods such as NN and GPR is an important technique for reducing the model-order complexity and run-time costs for inference. SSL models that are trained with randomized subsets of samples trade measurement and prediction costs for localization accuracy. Increasing the density of measurement samples over the spherical grid results in a linear increase to both NN classification computational cost and accuracy, a quadratic and cubic increase to respective GP inference and training computational costs, and a non-linear increase to GP localization accuracy. We show how GP-SSL models using small and non-uniform subset-selected samples (which are most informative) make more accurate predictions over the full spherical grid than models evidenced on a randomized subset. 

A simple greedy forward-selection (GFS) algorithm \cite{GUYON} that sequentially incorporates training samples into a subset without considerations in future iterations is implemented. It ranks all training samples outside the subset via a user-defined objective function (risk function) and adds the minimizer into the subset. We propose a class of risk functions that generalizes the GP prediction errors and show that the subset-selected GP-SSL models localize directions more accurately than models evidenced on randomized inputs (see section \ref{SEC:GPAL:GFS}); only a small fraction of training samples are required for reasonable accuracy ($5^{\circ}$).

\subsection{Active-learning for individualizing HRTFs}
 
Individualized HRTFs are needed for synthesizing accurate spatial audio that resolve front-back and up-down directional confusion \cite{KULKARNI1998,ROMIGH,WENZEL2}. Due to the difficulties of directly measuring HRTFs \cite{ZOTKIN2003}, a number of works have sought indirect means for learning the subject's HRTFs:  regression models between the individual's physically measured anthropometry and his/her HRTFs can be learned via neural-network \cite{HU} and multiple non-linear regression models \cite{HUANG} but do not generalize well to test subjects. HRTFs can also be learned through listening tests \cite{FINK,SILZLE} by having an individual listen to a query HRTF $\textbf{x}$ convolved with white Gaussian noise (WGN) (heard over a pair of headphones), localize the test signal (report a direction $\textbf{v}  \in \field{R}^{3}$), and then hand-tune the spectra of $\textbf{x}$ or choose a new $\textbf{x}$ out of a large candidate pool over a graphical user interface (GUI) as to move subsequent localizations towards a target direction $\textbf{u}  \in \field{R}^{3}$. The hand-tuning/selection step can be replaced by developing a recommendation system that selects for the query HRTF between rounds (steps) of localization.  The listener can rank candidate HRTFs chosen from a genetic algorithm\footnote{Evaluates a fitness function w.r.t. localization accuracy of known $\textbf{u}$} \cite{RUNKLE}. HRTFs can also be tuned along a low-dimensional autoencoder space \cite{LUO6} where $\textbf{u}$ is unknown to the listener.

We propose to formulate the recommendation problem in an \emph{active-learning} \cite{SETTLES} context described as follows: given a finite set of candidate HRTFs $X^C$ sampled from a prior distribution (database or generative model), determine the HRTF from the $X^C$ that the listener would localize nearest to $\textbf{u}$ within $T$ rounds of localizations. During round $t \leq T$, the recommender selects a query $\textbf{x}$  that the listener labels as $\textbf{v}_t(\textbf{x})$ without knowledge of $\textbf{u}$. The choice of $\textbf{x}$ is referred to as the \emph{query-selection} problem of minimizing the SSL error (SSLE) (modified cosine distance) given by
\begin{equation}
\label{EQ:GPAL:INTRO:SSLE}
\displaystyle
\begin{split}
\textbf{SSLE}(\textbf{u}, \textbf{v}_t(\textbf{x})) = -\textbf{u}^T \textbf{v}_t(\textbf{x}), \quad 
\arg \min_{\textbf{x} \in X^C} \textbf{SSLE}\PAREN{\textbf{u}, \textbf{v}_t(\textbf{x})}.
\end{split}
\end{equation}

Unfortunately, the minimizer in Eq. \ref{EQ:GPAL:INTRO:SSLE} is unlikely to be found within $T$ rounds as $X^C$ can be large and $T$ must also be small as the cost of evaluating SSLE by the listener is high. It is more reasonable to model the SSLE function using an online regression model (adapting HRTFs predictors of SSLEs after each round) and select for $\textbf{x}$ based on two competing strategies: query-selection  \emph{exploits} the online model by choosing $\textbf{x}$ that the model predicts will have low SSLE and \emph{explores} $\textbf{x}$ that has high model uncertainty in its prediction; both concepts are trade-offs that require probabilistic treatments of model predictions. Fortunately, GPs are well-suited to this task as all predictions are expressed as probabilistic realizations sampled from normal distributions. Thus, we propose to solve the modeling problem via \textbf{GP-SSLEs}\footnote{GPs that predict the SSLE from HRTFs}, and the query-selection problem using a method of GPs for the global optimization of smooth functions \cite{OSBORNE,SRINIVAS} (see section \ref{SEC:GPAL:ALM}). The relation between these methods and the GP-SSL models is also shown.

\section{Binaural Sound-Source Invariant Features}
\label{SEC:GPAL:FEATS}

\begin{figure}[ht]
  \centering
\includegraphics[width=.24\textwidth]{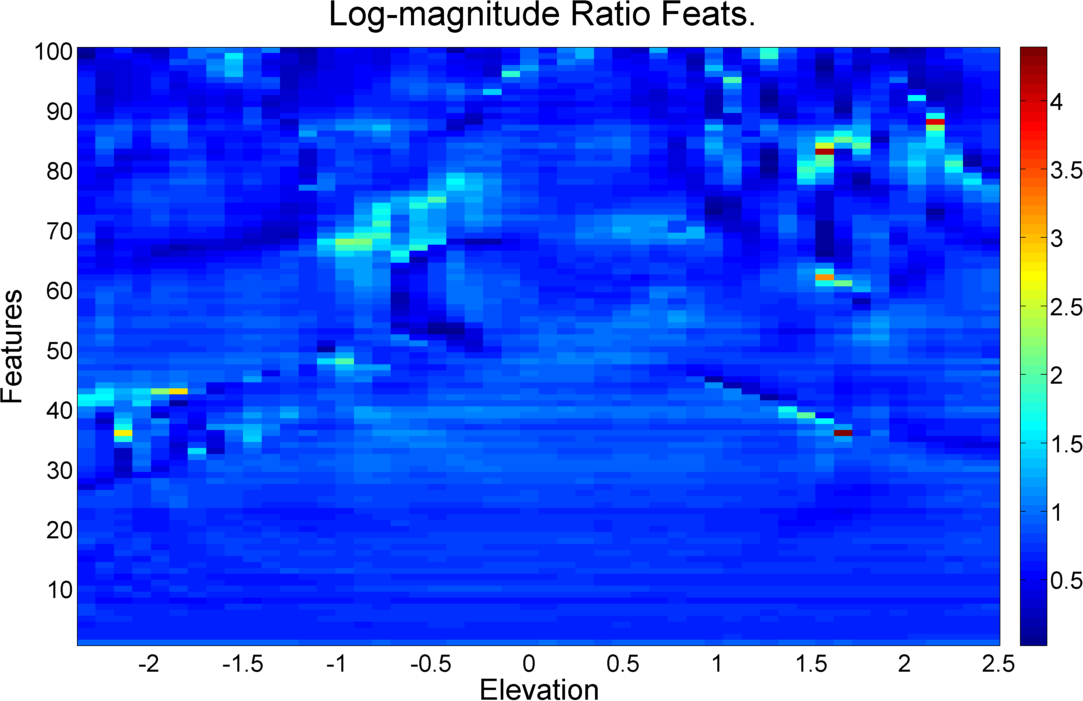}
\includegraphics[width=.24\textwidth]{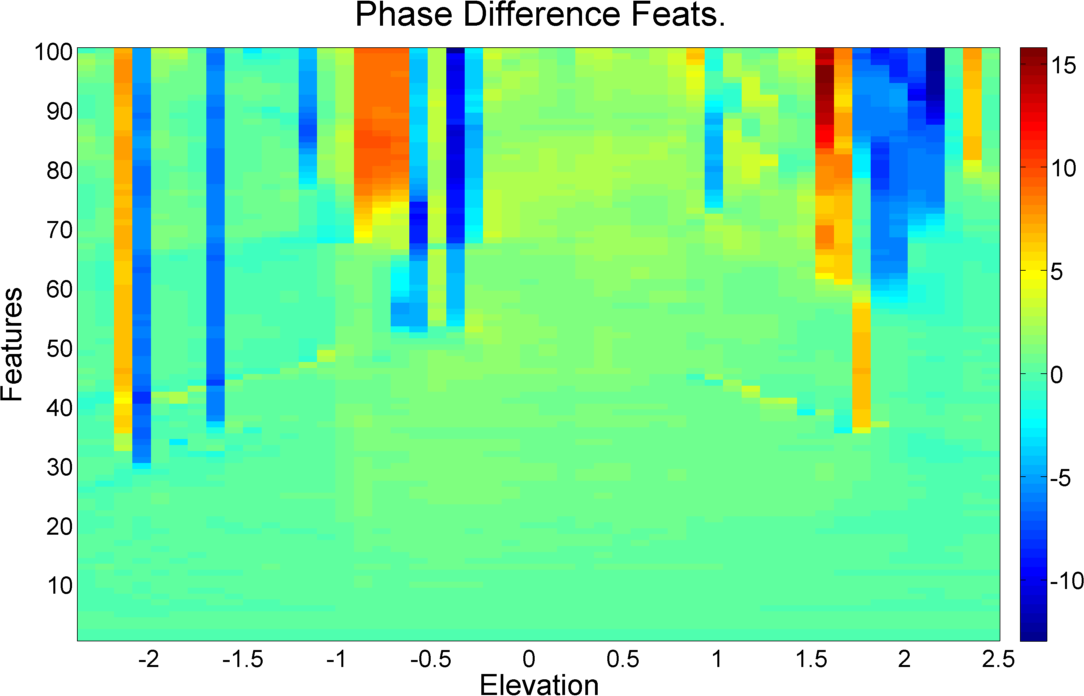}  \\ \vspace{.05cm}
\includegraphics[width=.24\textwidth]{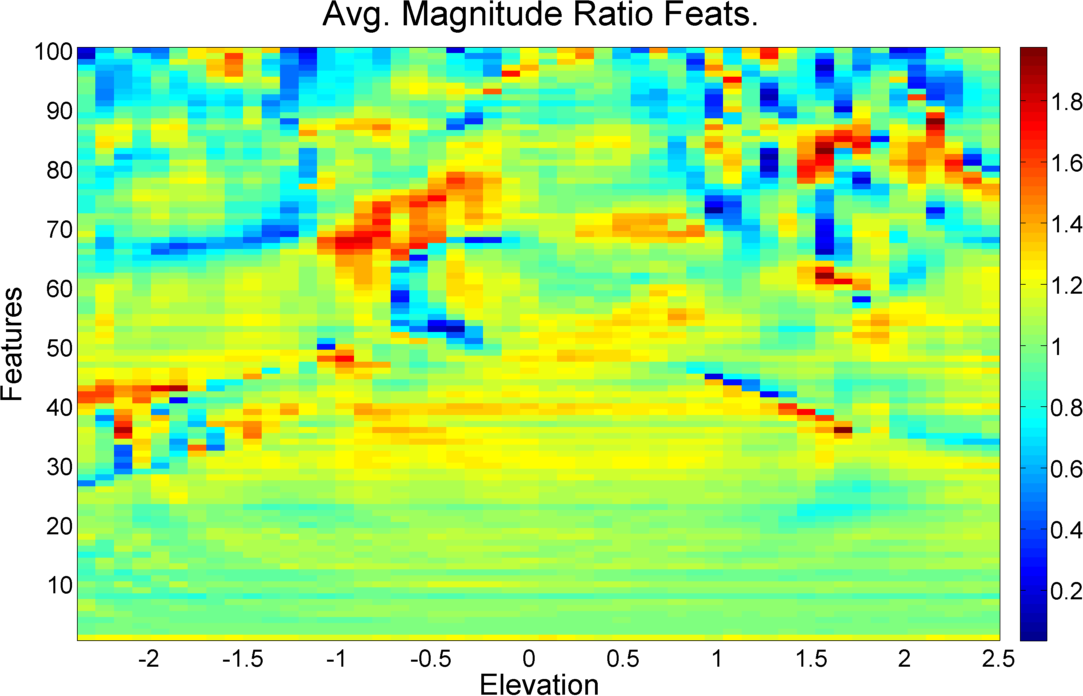}
\includegraphics[width=.24\textwidth]{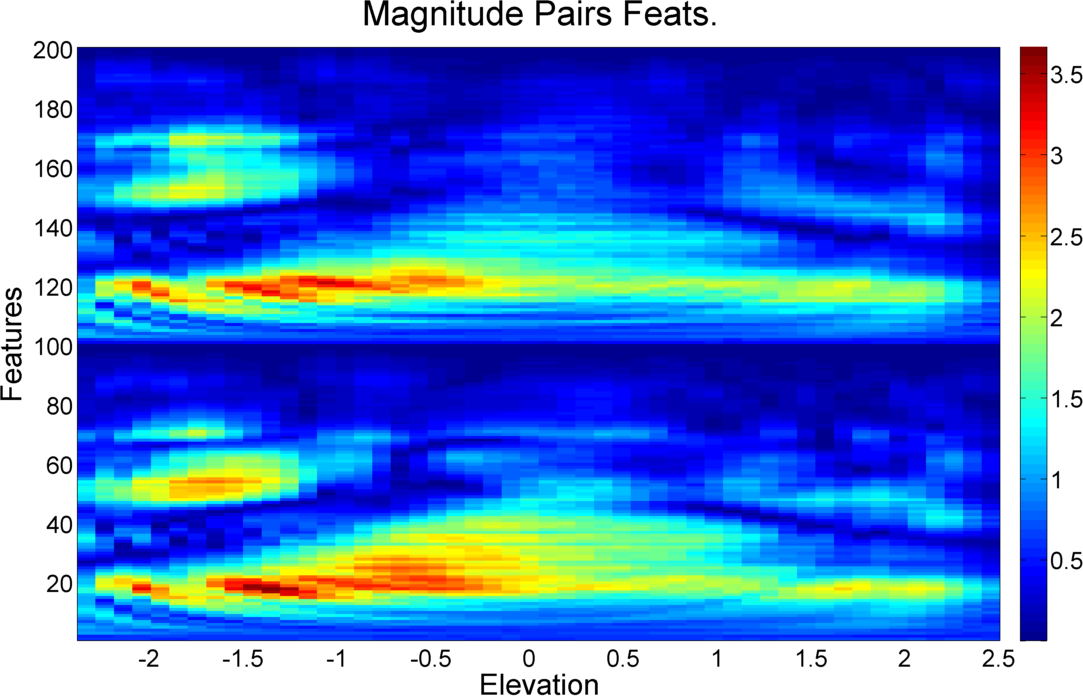}\\ \vspace{.05cm}
\includegraphics[width=.24\textwidth]{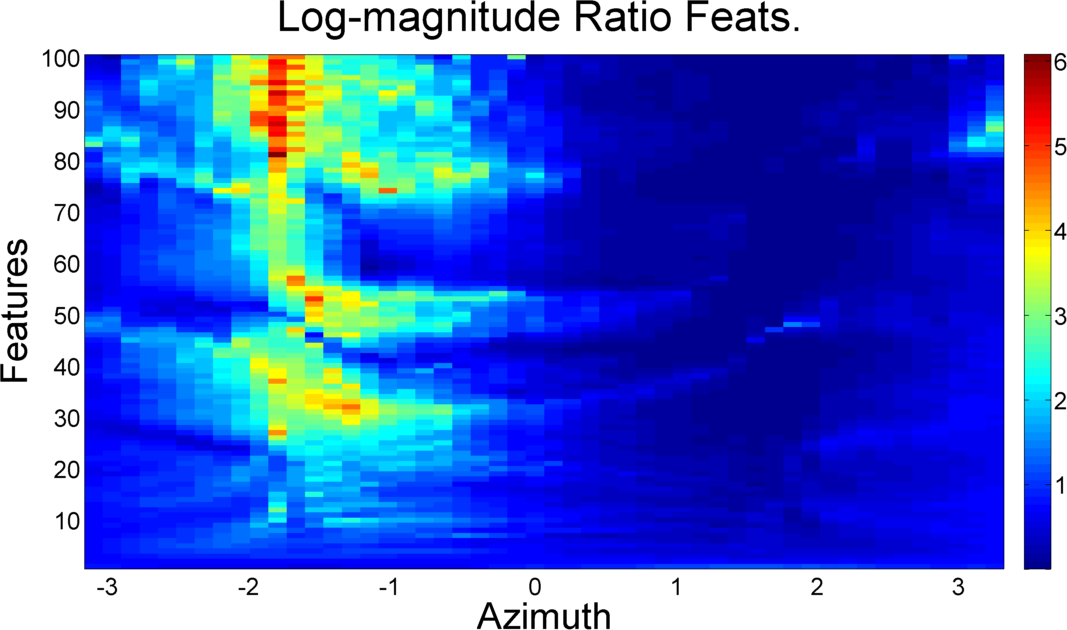}
\includegraphics[width=.24\textwidth]{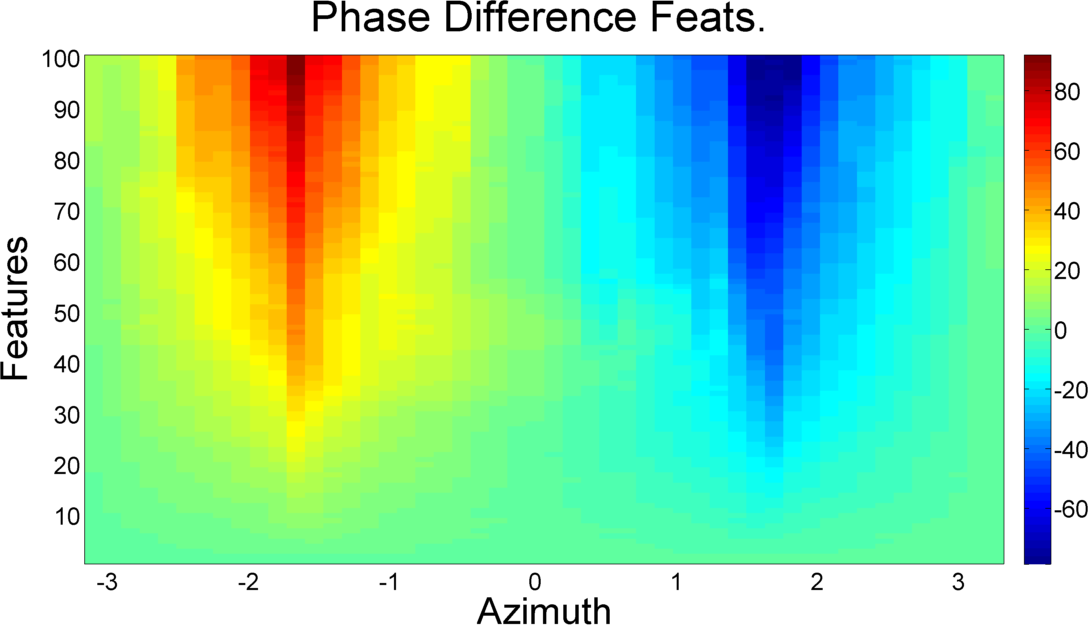}\\ \vspace{.05cm}
\includegraphics[width=.24\textwidth]{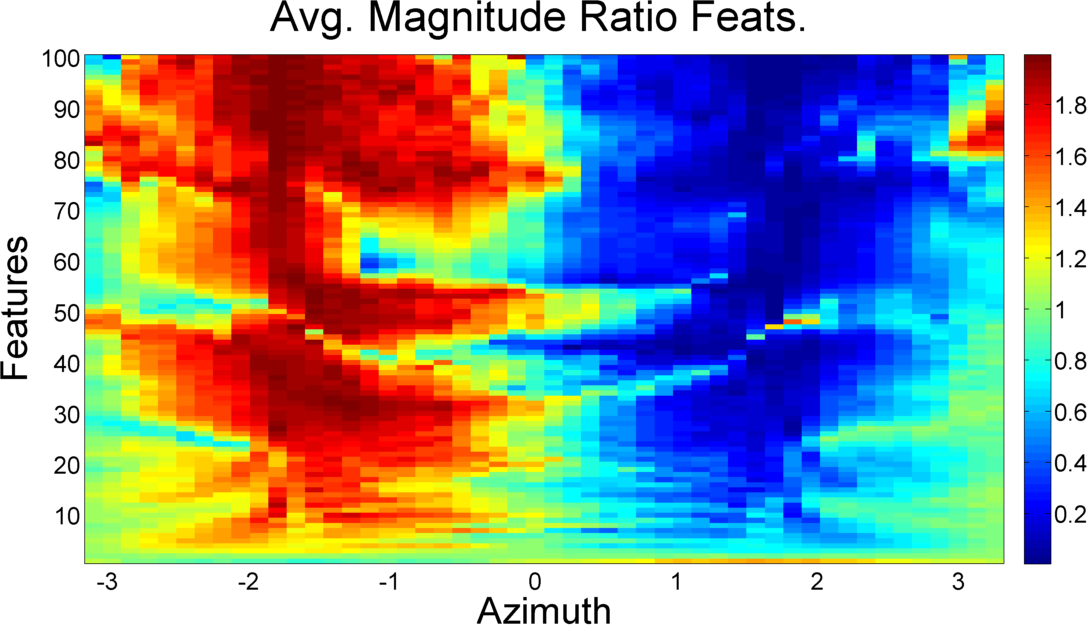}
\includegraphics[width=.24\textwidth]{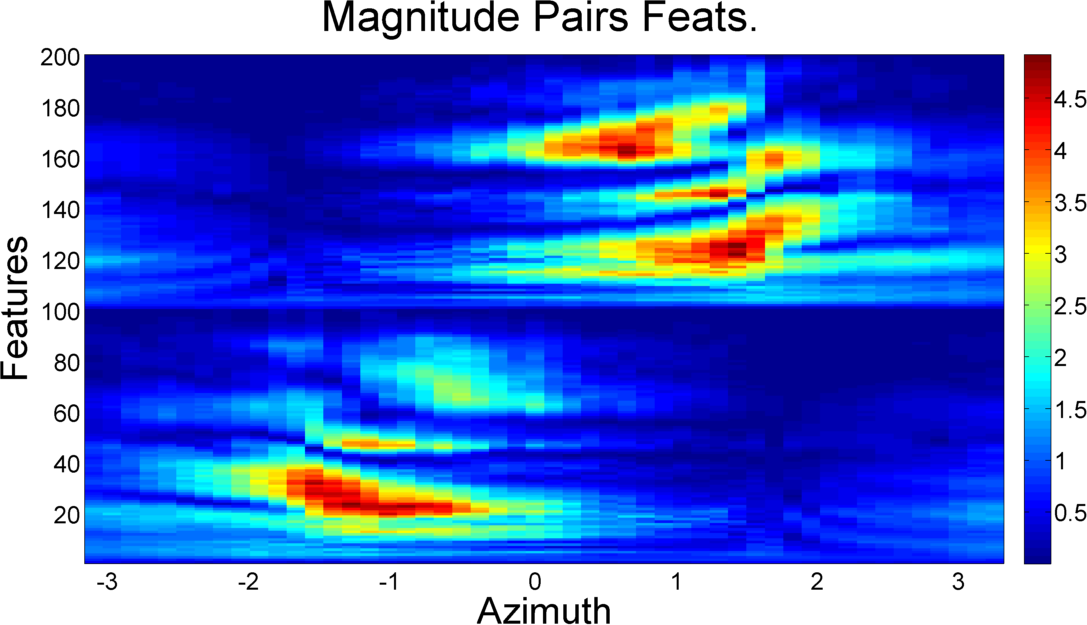}
\caption{Binaural features extracted from CIPIC subject $156$ HRTFs are shown for horizontal and median plane directions.
}
\label{FIG:GPAL:FEATS:ELEVNAZ}
\end{figure}

We consider several sound-source invariant features that can be extracted from short-time Fourier transforms of the left and right ear input channel streams in Eq. \ref{EQ:GPAL:INTRO:FREQCONV} (see Table \ref{TAB:GPAL:FEATS} and Fig. \ref{FIG:GPAL:FEATS:ELEVNAZ}); it is useful to express the discrete Fourier transformed signals by their magnitude and phase representations where $H(j\omega) = \ABS{H(j\omega)} e^{j \angle H(j \omega)}$. The features are expressed as ratios between left and right ear channel recordings that remove the effects of the acoustic content in $S$; the remainder is strictly a per-frequency function of same-direction left and right ear HRTFs derived as follows:

\begin{table}[ht]
\centering
\caption{HRTF sound-source invariant features $X$} \label{TAB:GPAL:FEATS}
\begin{tabular}{| c | l |} 
\hline
$\log\PAREN{\ABS{\frac{S_L}{S_R}}+1} = \log \PAREN{\frac{\ABS{H_L}} {\ABS{H_R}}+1}$ &  Log-magnitude ratio\\ \hline
$\angle \frac{S_L}{S_R} =  \angle H_L - \angle H_R$ &  Phase difference\\ \hline
$\frac{\ABS{S_L}}{0.5\PAREN{\ABS{S_L} + \ABS{S_R}}} = \frac{2\ABS{H_L}} {\ABS{H_L} + \ABS{H_R}}$ & Avg. magnitude ratio\\ \hline
$\CBRAK{\ABS{S_L}, \ABS{S_R}}= \CBRAK{\ABS{H_L}, \ABS{H_R}}$ & Magnitude pairs for flat $S$ \\ \hline
\end{tabular}
\end{table}


\textbf{Log-magnitude ratio (LMR) \cite{KEYROUZ1}:}
While the source-cancellation method removes the dependence on signal $S$, the resulting features are complex, noisy, and difficult to interpret. This can be avoided by considering the magnitude representation which gives the relative per-frequency energy between the channel signals. Adding a constant to the ratio prior to the log-transform penalizes the magnitude of the perturbation; adding a constant $1$ constrains the log-transform to be non-negative.

\textbf{Phase difference (PD):}
Similarly, the per-frequency phase of the complex channel signal ratio can be expressed by the phase-difference between left and right HRTFs. For identical $S_L, S_R$ that differ by onset time-delays $\Delta_L, \Delta_R$, the phase-difference is simply the constant delay $\Delta_L - \Delta_R$ across all frequencies; this ITD can be related to azimuth angles via Woodworth's model \cite{WOODWORTH}. For arbitrary $S_L, S_R$, the per-frequency phase-differences differ and are to be treated as independent variables in regression models.

\textbf{Average magnitude ratio (AMR):}
The magnitude source-signal $|S|$ can also be removed by taking the ratio of left or right magnitude signals $|S_L|,|S_R|$ and the binaural average $(|S_L| + |S_R|)/2$. Without the constant factor, the feature can be interpreted as the per-frequency contribution of the left or right magnitude HRTFs to the additive binaural magnitude response. Unlike log-magnitude ratio features that approaches a singularity as $|H_R| \rightarrow 0$, these features are bounded in the interval $[0, 2)$ and finite everywhere unless the binaural average is zero. 

\textbf{Magnitude pairs (MP):} 
The magnitude pairs are the concatenation of the original left and right magnitude HRTFs that could be derived from convolution with a WGN $S$ with zero mean and unit variance. The power spectrum of $|S|^2$ is constant across all frequencies and so $|S_L|, |S_R|$ would be constant factors of magnitude HRTFs. Such conditions arise during listening tests where the source-signal $S$ can be specified; the test features can then be derived from per-frequency division given by $H_L = S_L / S$ and $H_R = S_R / S$.

\section{Gaussian Process Regression for SSL}
\label{SEC:GPAL:GPR}

In a general regression problem, one predicts a scalar target variable $y$ from an input vector $\textbf{x}$ of independent variables based on a collection of available observations. A common Bayesian approach for inference assumes that the observation $y$ is generated (realized) from a latent function $f(\textbf{x})$ given by
\begin{equation}  \label{EQ:GPAL:GPR:LATENT}
\displaystyle 
\begin{split}
y & = f(\textbf{x}) + \epsilon, \quad \epsilon \sim \mathcal{N} (0, \sigma^2),
\end{split}
\end{equation}
which is corrupted by additive Gaussian white noise with zero mean and constant variance $\sigma^2$. This latent function is given the form of a kernel regression $f(\textbf{x}) = \phi(\textbf{x})^T \beta, \quad \beta \sim \mathcal{N}(0, \Sigma_p)$ where the function $\phi(\textbf{x}): \field{R}^D \rightarrow \field{R}^{D^*} $ maps the inputs $\textbf{x}$ into a high-dimensional space before computing the inner product with a vector of parameters realized from a collection of random variables with a prior multivariate normal distribution $\beta \in \field{R}^{D^*}$. Unlike linear regression, the parameters $\beta$ are not explicitly found in order to perform inference but are marginalized in order to compute the first two moments (mean and covariance) of function $f(\textbf{x})$ given by 
\begin{equation}  \label{EQ:GPAL:GPR:MOMENTS}
\displaystyle 
\begin{split}
\EX{f(\textbf{x})} & = \phi(\textbf{x})^T \EX{\beta } = 0, \\
\EX{f(\textbf{x})f(\textbf{x}')} & = \phi(\textbf{x})^T \EX{\beta \beta ^T} \phi(\textbf{x}') =  \phi(\textbf{x})^T \Sigma_p \phi(\textbf{x}').
\end{split}
\end{equation}
The latent function realizations $f(\textbf{x})$ are thus drawn from a multivariate normal distribution with mean $\mu(\textbf{x}) = 0$ and variance $k(\textbf{x},\textbf{x}') = \phi(\textbf{x})^T \Sigma_p \phi(\textbf{x}')$. For $\Sigma_p=I$, the inner product can be replaced with the covariance function $k(\textbf{x},\textbf{x}') =  \phi(\textbf{x})^T  \phi(\textbf{x}')$ which GPs generalize as follows:

A GP $f$ is a collection of random variables where any finite subset indexed at $N$ inputs $X=[\textbf{x}_1, \hdots,  \textbf{x}_N]$ has the joint multivariate normal distribution given by
\begin{equation}  \label{EQ:GPAL:GPR:GPPRIOR}
\displaystyle 
\begin{split}
[f(\textbf{x}_1), \hdots, f(\textbf{x}_N)] \sim \mathcal{N}(\mu(X), K(X,X)),
\end{split}
\end{equation}
and thus fully defined by the prior mean function $\mu(\textbf{x})$ and the prior covariance function $k(\textbf{x},\textbf{x}')$. The prior mean function and vector $\mu(X) \in \field{R}^{N}$ are set to zero without loss of generality following Eq. \ref{EQ:GPAL:GPR:MOMENTS}. The covariance (Gram) matrix $K(X,X) \in \field{R}^{N \times N}$ is characterized by the pairwise covariance function evaluations $K_{ij} = k(\textbf{x}_i, \textbf{x}_j)$; the covariance function is a positive semi-definite kernel (Mercer's condition) that establishes the existence of the eigenfunction $\phi(\textbf{x})$. This allows kernel methods such as SVR and GPR to omit computing the exact mapping $\phi$ as the inner products in the high-dimensional space, representing the similarity measure between input features $\textbf{x}, \textbf{x}'$, are well-defined.

GP inference at test inputs $X_* \in \field{R}^{N_* \times D}$ evidenced on training inputs $X$ and the observations in $Y \in \field{R}^{N}$ derives from the multivariate normal distribution of random variables $f_* = f(X_*)$ conditioned on $f(X)=Y$, $X$. This is given by
\begin{equation}  \label{EQ:GPAL:GPR:PREDICTION}
\displaystyle 
\begin{split}
f_* | X, Y, X_*  & \sim \mathcal{N} (\bar{f}_*, \bar{\Sigma}_*), \quad \bar{f}_*  =  K_{f*}^T \hat{K}^{-1} Y,\\
\bar{\Sigma}_*  & = K_{**} - K_{f*}^T \hat{K}^{-1} K_{f*},
\end{split}
\end{equation}
where $\hat{K} = K(X,X) + \sigma^2 I$ adjusts for the observation noise and $K_{f*} = K(X, X_*) \in \field{R}^{N \times N_*}$ are pair-wise covariance evaluations between training and test inputs. We refer to the distribution in Eq. \ref{EQ:GPAL:GPR:PREDICTION} as the posterior GP defined by the \emph{posterior mean} and \emph{posterior covariance} functions  $\bar{f}_*$ and $\bar{\Sigma}_*$ respectively. The former represents the vector  of expected outputs (prediction) at $X_*$ and the latter is gives the confidence intervals (diagonal of the matrix) of the predictions.

For the GP-SSL model, $X$ and $Y \in \field{R}^{N \times 3}$ are the respective binaural features in Table. \ref{TAB:GPAL:FEATS} and their measurement directions (unit vectors where $Y_i = Y_{:,i}$ are values along the $i^{th}$ coordinate); test inputs $X_*$ refer to the binaural features extracted from test signals. While it is possible to model all $M=3$ output coordinates as a collection of $M$ independent GPs $f_{1:M}(X) = \CBRAK{f_1(X), \hdots f_M(X)}$, a computationally cheaper alternative is to specify a common prior mean and covariance function shared by all GPs. Specifying a shared covariance model between GPs is reasonable as the original HRTFs are originally measured over the same physical topology of a human subject from a near-uniform spherical grid of directions. Thus for inference, we use three independent GPs, with shared priors, to model left-right, front-back, and top-down coordinate directions by either sampling from their posterior distribution or reporting their posterior means.

\subsection{Choice of Covariance Functions}
\label{SEC:GPAL:GPR:COV}

The ``smoothness/correlatedness'' of realizations of $f(X)$ for similar $X$ depends on the number of times that the covariance function is differentiable w.r.t. the input arguments. Consider the Mat\'{e}rn  class of covariance functions where each function has varying orders of differentiation. For $D$-dimensional inputs, we can specify the GP covariance function as the product of $D$-independent Mat\'{e}rn  covariance functions of identical class. Three common classes and the product covariance function are given as
\begin{equation} \label{EQ:GPAL:GPR:COV}
\displaystyle
\begin{split}
K_{\frac{1}{2}}(r, \ell) & = e^{-\frac{r}{\ell}}, \quad
K_{\frac{3}{2}}(r, \ell) = \PAREN{1+\frac{\sqrt{3}r}{\ell}} e^{-\frac{\sqrt{3}r}{\ell}}, \\
K_{\infty}(r, \ell) & = e^{-\frac{r^2}{2\ell^2}}, \quad K(\textbf{x},\textbf{x}') = \alpha^2 \prod_{k=1}^{D} K_{\nu}(|\textbf{x}_k-\textbf{x}_k'|, \ell_k),
\end{split}
\end{equation}
for distance $r$ and hyperparameters $\alpha, \ell_k$. Covariance functions $K_\nu$ are $\FLOOR{\nu}$ times differentiable and stationary due to their dependence on $\ABS{\textbf{x}_k-\textbf{x}_k'}$. Each function contains a length-scale or bandwidth hyperparameter $\ell_k$ that represents a distance in the domain $\textbf{x}_k$ where outputs $f(\textbf{x}_k)$ remain correlated; larger length-scales result in smoother $f$.

A general hyperparameter $\Theta$ is optimized by maximizing the data log-marginal likelihood (LMH) of the observations $Y$ given the GP prior distributions; the derivation follows from integrating over the realizations $f(X)$ by the product of data likelihoods (sampling $Y$ from $f(X) + \epsilon$ and sampling $f(X)$ from the GP prior distribution). The LMH term $L = \log p(Y|X)$ and its partial derivative are both analytic and given by
\begin{equation}  \label{EQ:GPAL:GPR:COV:LOGLIKEGRAD}
\displaystyle 
\begin{split}
L & = - \frac{M}{2} \PAREN{\log|\hat{K}| + \frac{\trace{Y^T \hat{K}^{-1} Y}}{M} + N
\log (2\pi) }, \\
\pderiv{L}{\Theta_i} & = - \frac{M}{2} \PAREN{\trace{  \hat{K}^{-1} P} - \frac{\trace{Y^T \hat{K}^{-1} P \hat{K}^{-1} Y} }{M}},
\end{split}
\end{equation}
where $P=\partial \hat{K} / \partial \Theta$ is the matrix of partial derivatives. A larger LMH represents a better goodness-of-fit of the data to the GP prior mean and covariances assumptions. Moreover, different covariance functions with optimized hyperparameters can be compared in this respect without resorting to domain-specific metrics.

\subsection{Model-Order and Cost Analysis}

The GP model-order is proportional to the size of the GP prior distribution defined by the $N$-dimensional multivariate normal distribution in Eq. \ref{EQ:GPAL:GPR:GPPRIOR} ($N$ is the number of training samples). The associated costs of both conditioning on the GP prior distribution for inference and performing hyperparameter training is dominated by the inversion of the Gram matrix ($\BIGO{N^3}$ operations to compute and $\BIGO{N^2}$ space to store). For large $N$, exact GP becomes intractable and most practitioners rely on randomized sampling techniques \cite{WILLIAMS} to reduce the costs at the expense of accuracy. Two types of analyses for evaluating this trade-off are given: first, empirical cross-validation experiments can demonstrate how data sampling (randomized and subset-selection) increases localization error. Second, the theoretical dimensionality of the feature space $\phi(\textbf{x})$ in Eq. \ref{EQ:GPAL:GPR:LATENT}, despite not having been explicitly computed, can be estimated from an eigenanalysis of the GP Gram matrix. The distribution of eigenvalues (number of dominant ones) gives a minimum bound as to the number of input features whose mapping will contain most of the variances in the feature space. 

To evaluate the dimensionality of $\phi(\textbf{x})$, we refer to the method of kernel principal component analysis \cite{MIKA} of Gram matrix $K$. Its derivation expresses the eigenvectors $v$ (principal directions) and eigenvalues $\lambda$ (measure of variance captured by $v$) of the sample covariance matrix $\tilde{C}$ of features $\phi(\textbf{x})$ in the high-dimensional space in the form of
\begin{equation} \label{EQ:GPAL:GPR:MODELORDER:KPCA}
\displaystyle 
\begin{split}
\tilde{C}  = \frac{1}{N} \sum_{i=1}^N \phi(\textbf{x}_i) \phi(\textbf{x}_i)^T, \quad
\tilde{C}v = \lambda v, \quad 
v = \frac{\sum_{i=1}^N \alpha_i \phi(\textbf{x}_i)}{\lambda N}, 
\end{split}
\end{equation}
where $\alpha_i = \phi(\textbf{x}_i)^T v$ are the component scores between the feature mapping and the eigenvector. Applying the ``kernel'' trick allows $\alpha$ to be reformulated in terms of the Gram matrix $K$ as a tractable eigendecomposition problem given by
\begin{equation}
\displaystyle 
\begin{split}
\sum_{j=1}^N \lambda \alpha_j & = \sum_{j=1}^N \phi(\textbf{x}_j)^T \tilde{C}v = \frac{1}{N}  \sum_{j=1}^N  \sum_{i=1}^N \alpha_j K_{ij}, \\
K_{ij} & = \phi(\textbf{x}_i)^T \phi(\textbf{x}_j), \quad K\alpha = \lambda N \alpha,
\end{split}
\end{equation}
which finds the eigenvalues $\lambda$ and scores $\alpha$. Evaluating the contributions of the leading $\lambda$ to the total energy $\sum_{i=1}^N \lambda_i$ estimates the number of eigenvectors that are relevant to $\phi(\textbf{x})$.

\subsection{Experiments}
GP-SSL models (input binaural features LMR, PD, AMR, and MP from Table. \ref{TAB:GPAL:FEATS} belonging to CIPIC subject $156$) are trained (batch gradient descent of all covariance function hyperparameters $\ell_k$ via Eq. \ref{EQ:GPAL:GPR:COV:LOGLIKEGRAD}) for $50$ iterations. For a domain-metric, we use the angular separation distance between two directions $\textbf{u, u'}$ (predicted and reference directions) given by
\begin{equation}\label{EQ:GPAL:GPR:EXP:AS}
\displaystyle
\begin{split}
\dist{\textbf{u}, \textbf{u'}} =  \cos^{-1} \frac{<\textbf{u}, \textbf{u'}>}{||\textbf{u}|| ||\textbf{u'}||}, \quad \textbf{u}, \textbf{u'} \in \field{R}^{3}.
\end{split}
\end{equation}

\textbf{Goodness-of-fit:}
GP-SSL models are specified/trained on the full set of inputs $X$. The data LMHs in Table. \ref{TAB:GPAL:GPR:EXP:LMH} are computed for several covariance functions and feature types. The infinitely differentiable squared exponential $K_{\infty}$ gives the best-fit (highest LMH) across all features (latent functions modeling the SSL directions are smooth w.r.t. changes in the feature space). This confirms the fact that a finite collection of HRTFs approximates a sound-pressure field that is continuous in space. The best-fitting binaural features are the MPs (WGN sound-source) and AMRs (arbitrary sound-source); the LMH gap between the two suggest that GP-SSL models will perform more accurately when the recorded magnitude spectra match that of the HRTFs. The LMH gap between AMR and LMR suggests that relative contribution may be a better indicator of SSL than relative intensities. The low LMH of PD models suggests that phase may not be useful for SSL over the entire spherical coordinate system.

\begin{table}
\centering
\caption{Data LMH for feature/GP covariance types} \label{TAB:GPAL:GPR:EXP:LMH}
\begin{tabular}{|c|c|c|c|c|}
\hline
&LMR&PD&AMR&MP\\\hline
$K_{\infty}$&\textbf{2.69e+003}&\textbf{2.37e+003}&\textbf{3.9e+003}&\textbf{6.34e+003}\\\hline
$K_{3/2}$&2.23e+003&1.5e+003&3.88e+003&6.29e+003\\\hline
$K_{1/2}$&2.06e+003&460&2.24e+003&4.84e+003\\\hline
\end{tabular}
\end{table}

\begin{figure}[ht]
  \centering
\includegraphics[width=.24\textwidth]{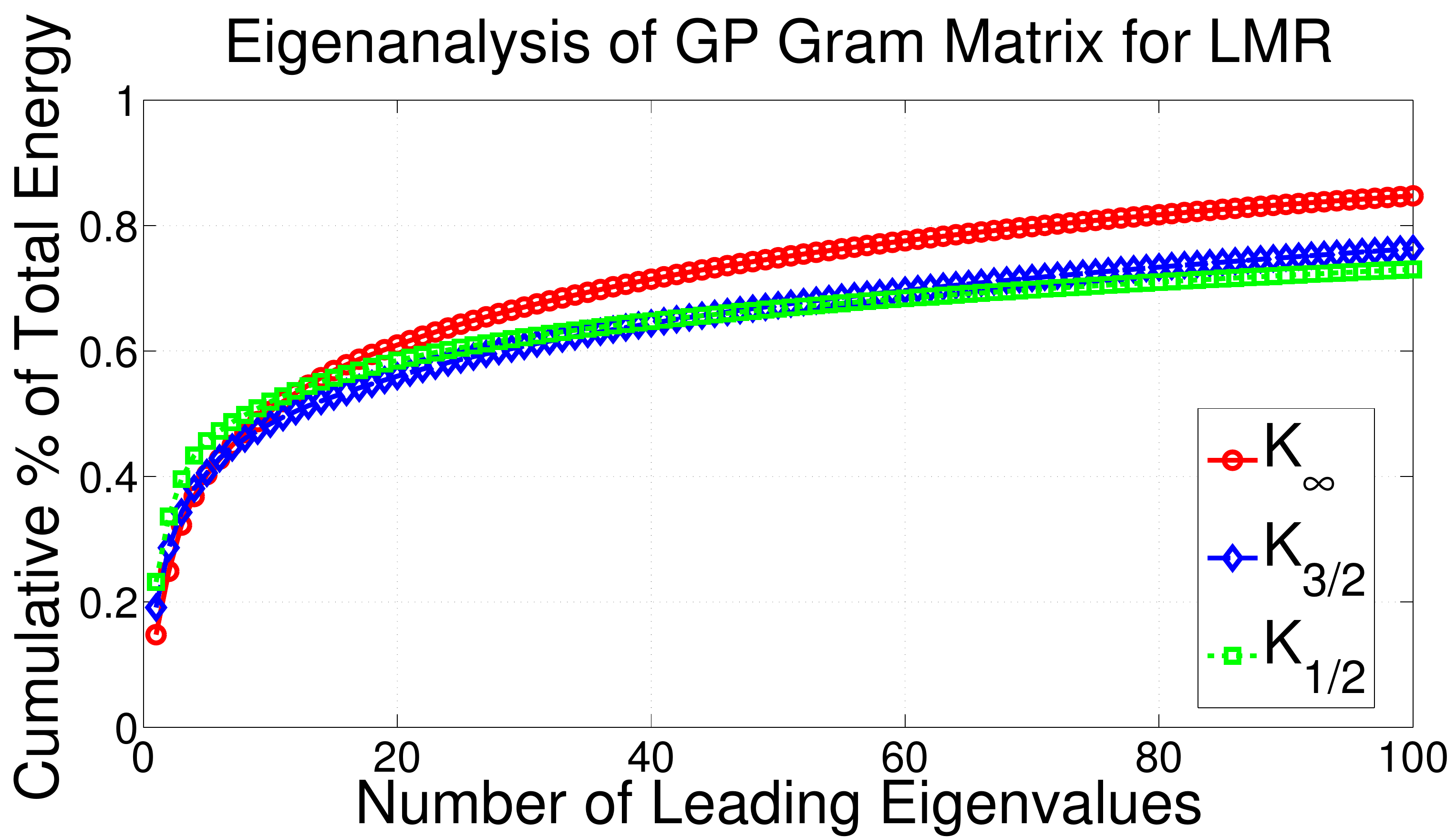}
\includegraphics[width=.24\textwidth]{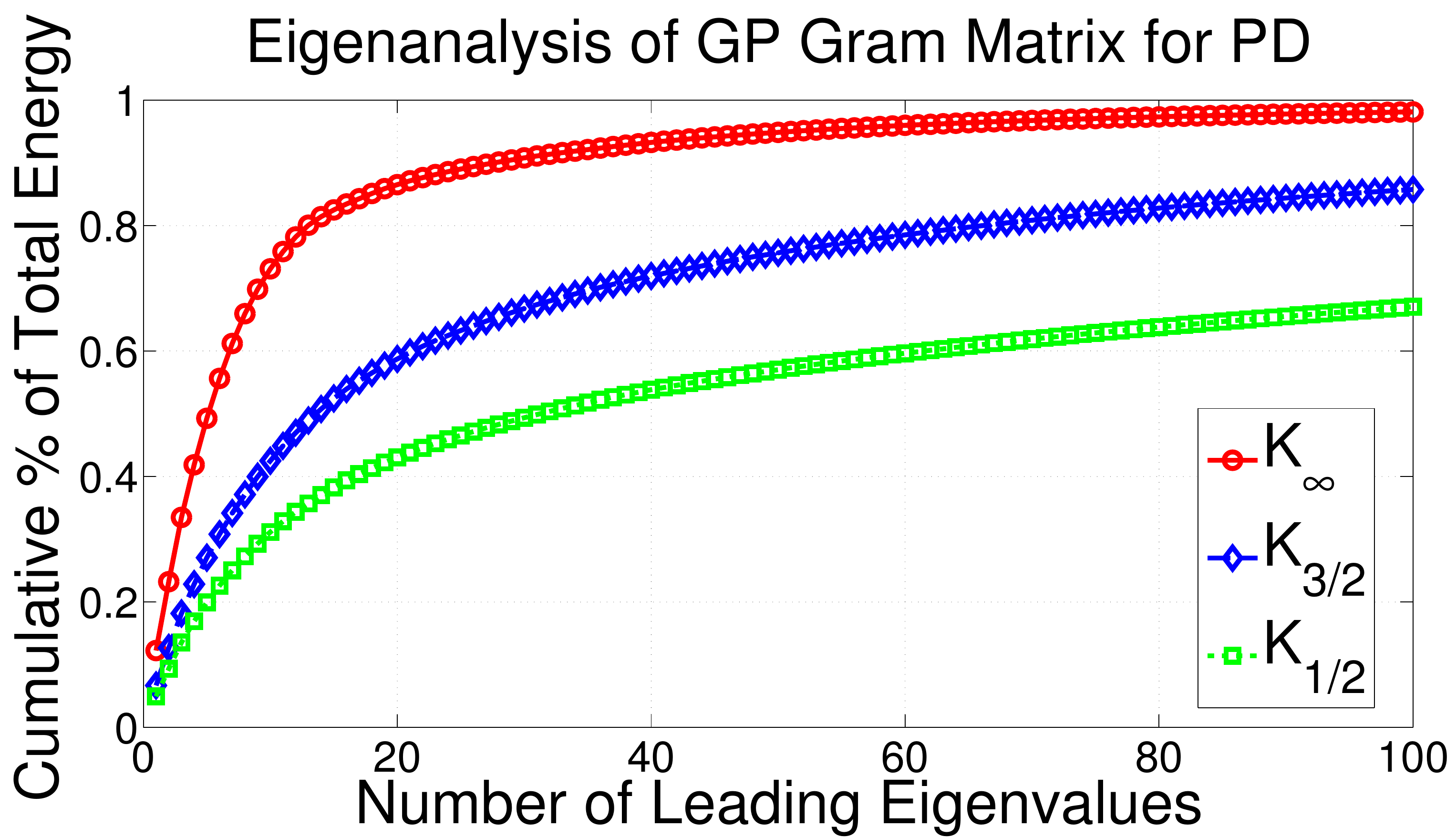}
\includegraphics[width=.24\textwidth]{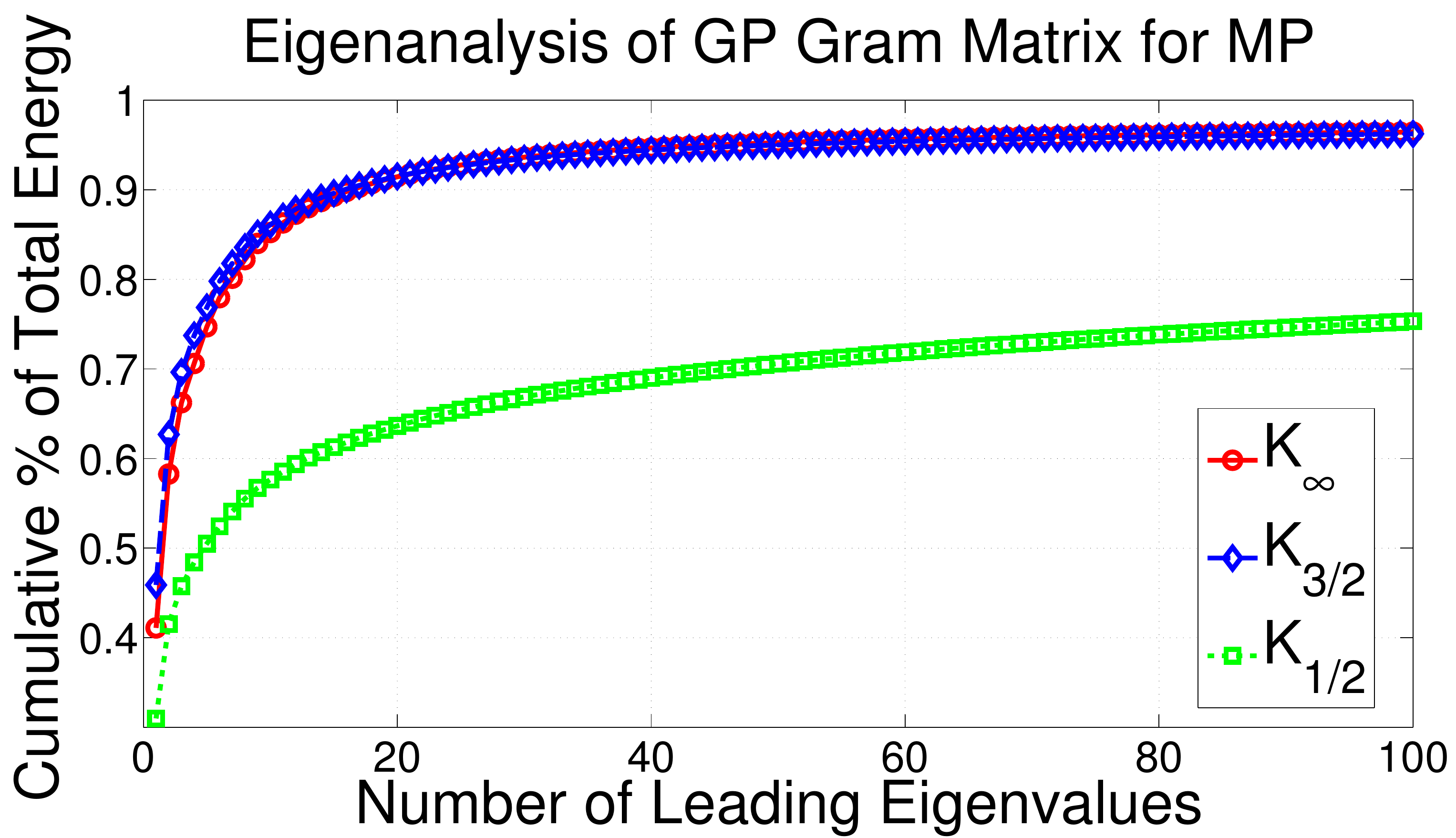}
\includegraphics[width=.24\textwidth]{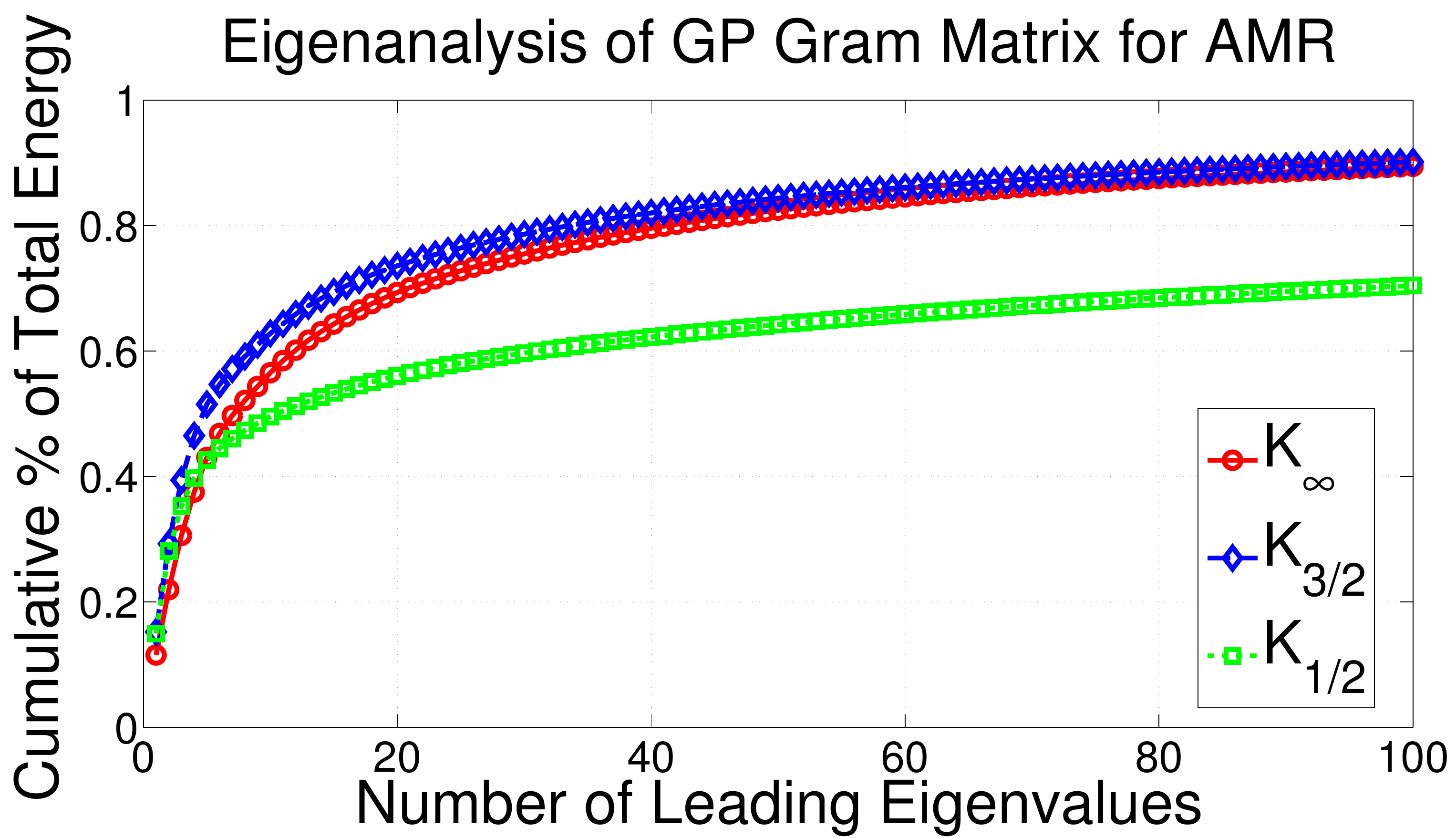}
\caption{Cumulative energy of leading eigenvalues for $K$ are shown for GP-SSL models (varying covariance functions and feature types).
}
\label{FIG:GPAL:GPR:EXP:EIGEN}
\end{figure}

\textbf{Eigenanaylsis of $K$:}
The eigenvalues of the $K$ are computed for GP-SSL models trained/specified on the full dataset ($N=1250$). Fig. \ref{FIG:GPAL:GPR:EXP:EIGEN} shows the contribution of the leading eigenvalues to the total energy; $K_{\infty}$ specified by the four earlier features (LMR, PD, AMR, and MP) require respectively $150$, $30$, $100$, and $15$ leading eigenvectors to capture $90\%$ of the total variance. The results suggest that feature mappings for MPs and PDs can be approximated with only a few samples while LMR and AMR feature mappings are more complex.

\begin{figure*}[ht]
  \centering
\includegraphics[width=.49\textwidth]{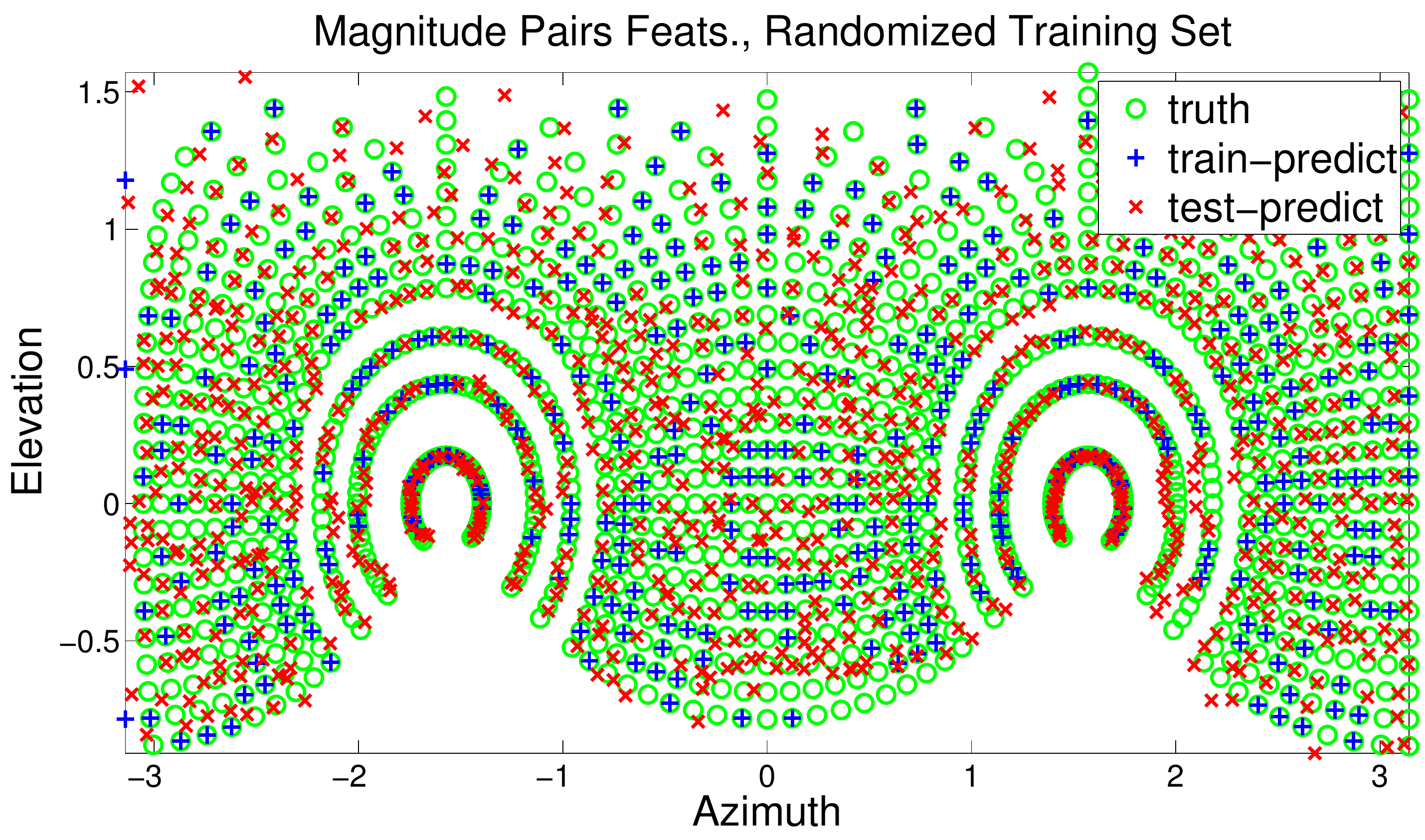}
\includegraphics[width=.49\textwidth]{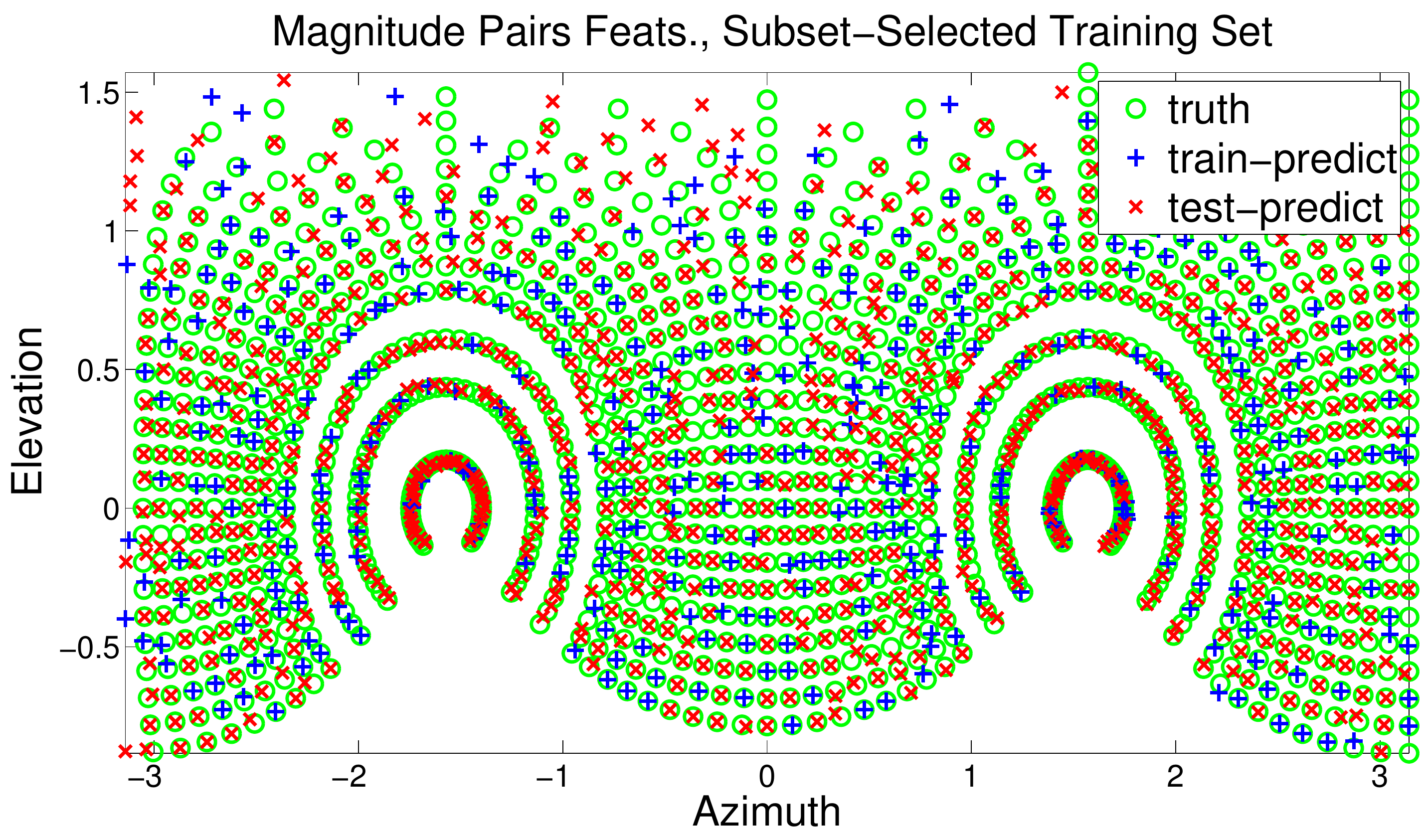} \vspace{.05cm}
\includegraphics[width=.49\textwidth]{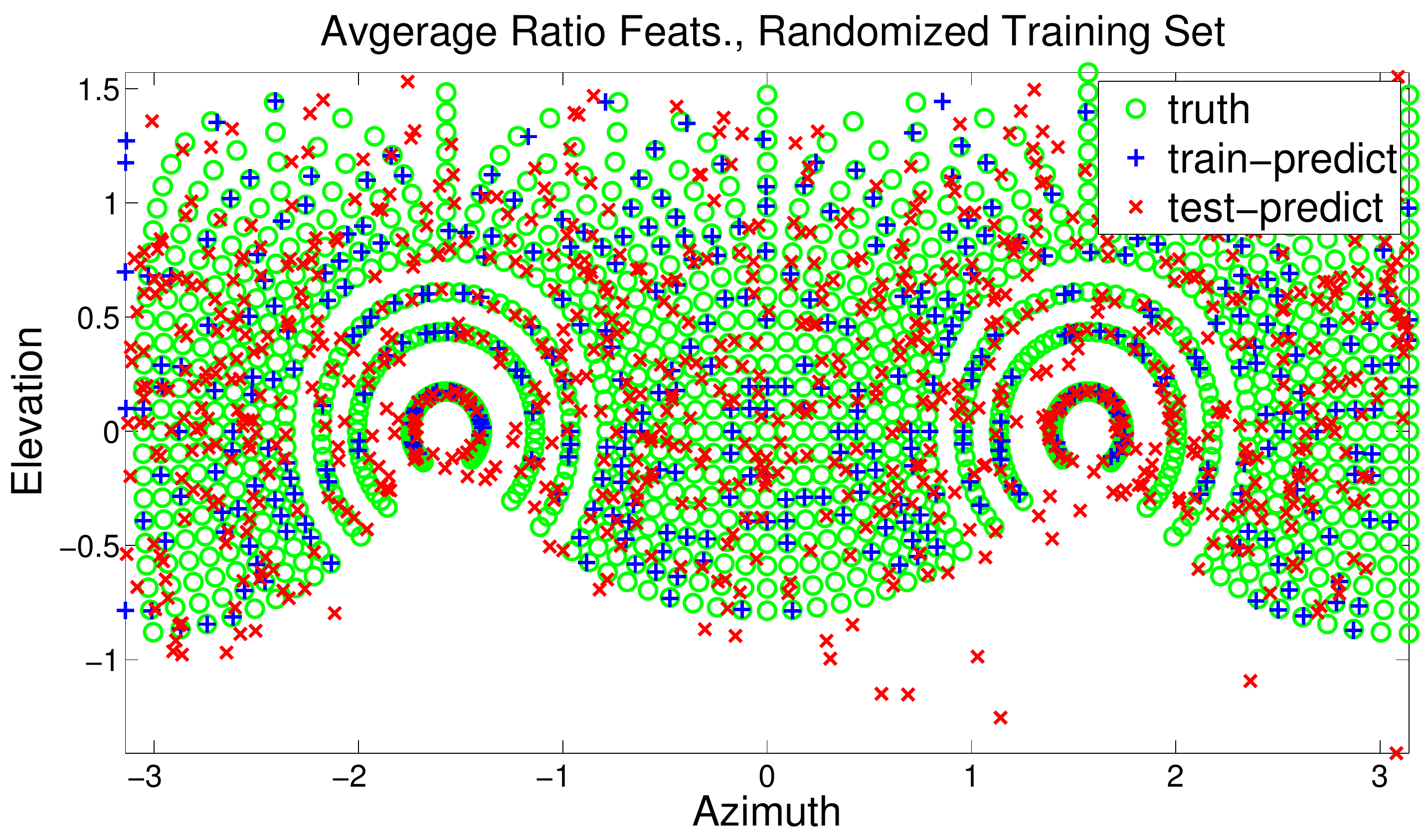}
\includegraphics[width=.49\textwidth]{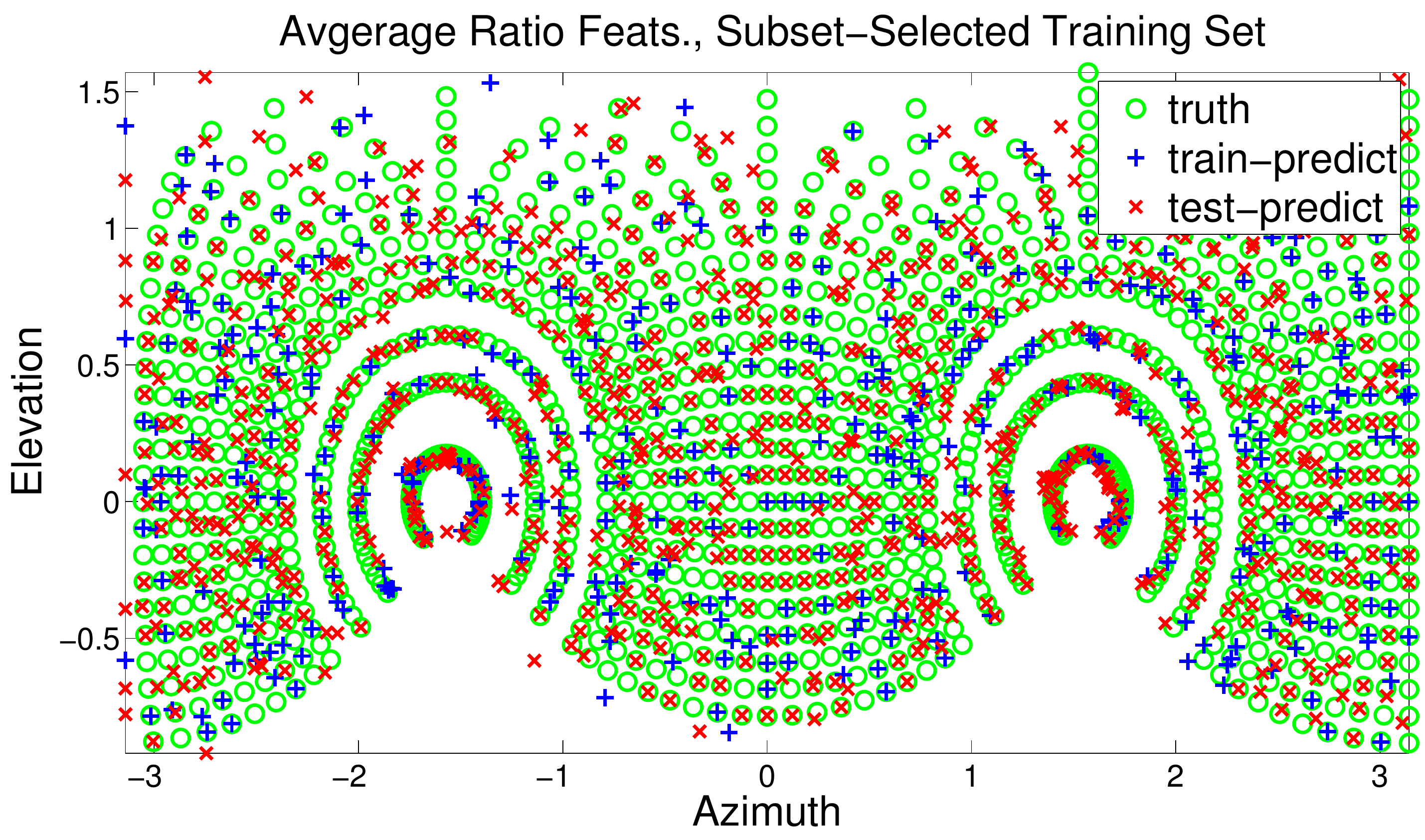}
\caption{Mercator projections of GP-SSL $K_{\infty}$ predicted mean directions evidenced on randomized and subset-selected inputs (prediction error risk function $R$ in section \ref{SEC:GPAL:GFS:RISK} are shown. 
}\label{FIG:GPAL:GPR:EXP:GEN}
\end{figure*}

\textbf{Cross-validation:} 
GP-SSL models are trained on a randomized third of the available feature-direction pairs ($N=417$ out of $1250$); inference follows Eq. \ref{EQ:GPAL:GPR:PREDICTION} at all available inputs ($X_* = X$) where only the posterior mean directions are reported. Table \ref{TAB:GPAL:GPR:EXP:GEN} shows the mean angular separation (Eq. \ref{EQ:GPAL:GPR:EXP:AS}) between predicted and reference directions for GP-SSL, NN classifier, OLS methods trained on the same data. Non-parametric methods (NN and GPR) outperform parametric methods (OLS) across all feature types. The MP and AMR features give the lowest errors across all methods (for a visual, see the first column of Fig. \ref{FIG:GPAL:GPR:EXP:GEN}). OLS log-ratios perform the worse and suggest that the features are oversensitive linear predictors of change in localization direction. PD features, while useful for predictions on the horizontal plane, are insufficient for localizations over the full sphere. 

\begin{table}
\centering
\caption{Mean angular separation errors (degrees) for feature/methods} 
\label{TAB:GPAL:GPR:EXP:GEN}
\footnotesize
\begin{tabular}{|c|c|c|c|c|}
\hline
&LMR&PD&AMR&MP\\\hline
OLS&29&27&22&5.4\\\hline
NN&9.2&20&7.9&3.9\\\hline
GP-SSL $K_{1/2}$&7.2&12&7&1.8\\\hline
GP-SSL $K_{3/2}$&7.5&11&4.8&1.4\\\hline
GP-SSL $K_{\infty}$&\textbf{6.3}&\textbf{6.3}&\textbf{4.8}&\textbf{1.3}\\\hline
\end{tabular}
\end{table}

\section{Feature Subset-Selection}
\label{SEC:GPAL:GFS}

Greedy feature selection is an efficient method for finding a subset of inputs $X_{r} \in X$ that best approximates a functional $f(X_{r}) \approx f(X)$ according to a user-specified risk function $R(X_{r})$ (measure of distance between $f(X_{r})$ and $f(X)$). Determining the optimal subset via an combinatorial exhaustive search is prohibitive w.r.t. the number of evaluations of $R$. A greedy heuristic (ranking $X_{\hat{r} \not \in r}$ according to a point-inclusion in the risk evaluation $R(X_{\hat{r} \cup r})$ and adding the minimizer into the subset $X_{r}$ without consideration in future iterations) reduces the search to a quadratic number of evaluations (see Algorithm \ref{ALG:GPAL:GFS}). For GP-SSL, GFS approximates the GP posterior distribution (Eq. \ref{EQ:GPAL:GPR:PREDICTION}) evaluated on the full dataset ($X_* = X$) conditioned on a growing subset $X_{\hat{r} \cup r}$ of inputs. We propose an efficient method for updating both GP prior and posterior distributions between point-inclusions in section \ref{SEC:GPAL:GFS:IGPR}. 

\begin{algorithm}
\caption{Greedy Forward Selection}
\label{ALG:GPAL:GFS}
\begin{algorithmic}[1]
\REQUIRE  Training inputs $X, y$, subset size $T$, and risk function $R(X)$.
\STATE $r \gets \emptyset$ \ALGCMT{Initial empty subset at iteration $t=0$}
\FOR {$t = 1$ to $T$}
	\STATE 	$r \gets \CBRAK{r, \arg \min_{\hat{r} \not \in r}{R\PAREN{X_{\hat{r} \cup r }}}}$ \ALGCMT{Minimize risk}
\ENDFOR
\RETURN $r$
\end{algorithmic}
\end{algorithm}

Specifying the risk function $R$ is more difficult as its evaluation costs must be low. Most risk functions that use second-order moments (e.g. GP posterior covariance in Eq. \ref{EQ:GPAL:GPR:PREDICTION}) are expensive and require approximations to remain tractable \cite{SEEGER}. Evaluating the GP posterior covariance requires $\BIGO{N_*^2}$ space; its inverse and determinants are expensive to compute in sub-cubic time. Instead, we propose a cheaper class of risk functions that generalizes only the first-order moments (i.e. GP posterior mean in Eq. \ref{EQ:GPAL:GPR:PREDICTION}) in section \ref{SEC:GPAL:GFS:RISK}.

\subsection{Incremental GP Models}
\label{SEC:GPAL:GFS:IGPR}

A point-update to a GP model can be defined in terms of changes to the first/second moments of the GP prior and posterior distributions (Eqs. \ref{EQ:GPAL:GPR:GPPRIOR}, \ref{EQ:GPAL:GPR:PREDICTION}) and both the Gram matrix $K_{\PAREN{r}} = K(X_{r}, X_{r})$ and its inverse $K_{\PAREN{r}}^{-1}$ generated from inputs in $X_{r}$.  While a point-update to $K_{\PAREN{ \hat{r} \cup  r}}$ simply contains an appended row and column of covariance function evaluations $\BRAK{K(X_{r}, x_{\hat{r}}), K(x_{\hat{r}}, x_{\hat{r}})}$, its direct inverse  $K_{\PAREN{\hat{r} \cup r}}^{-1}$ would be expensive to compute. Instead, we define a recurrence relation with its previous inverse $K_{\PAREN{r}}^{-1}$ as follows.

Given a sample input-output pair $(\textbf{x}_{\hat{r}}, \textbf{y}_{\hat{r}})$ for data index $\hat{r}$, let indices $\breve{r} = r \cup \hat{r}$ be the union with the subset indices $r$. At iteration $t$, append a row and column vector along the standard basis to the Gram matrix $K_{(r)}$. The differences between $K_{\PAREN{\breve{r}}}$ and the appended $K_{(r)}$ are two rank-$1$ updates given by
\begin{equation}\label{EQ:GPAL:GFS:IGPR:UP}
\displaystyle
\begin{split}
K_{\PAREN{\breve{r}}}  & = \BRAK{
\begin{array}{cc}
K_{(r)} & k_{r\hat{r}} \\
k_{r\hat{r}}^T &  k_{\hat{r}\hat{r}}
\end{array}} =  \BRAK{
\begin{array}{cc}
K_{(r)} & 0 \\
0 & 1
\end{array}}
 - uu^T + vv^T,\\
k_{r\hat{r}} & =  K(X_{r}, X_{\hat{r}}), \quad k_{\hat{r}\hat{r}} = K(X_{\hat{r}}, X_{\hat{r}}) + \sigma^2,
\end{split}
\end{equation}
where vectors $u  = \sqrt{\frac{|| w ||}{2}} \left ( \frac{w}{|| w || } + e_t \right )$, $v  = \sqrt{\frac{|| w ||}{2}} \left ( \frac{w}{|| w || } - e_t \right )$, $w  = \BRAK{-k_{r\hat{r}}^T, \frac{1-k_{\hat{r}\hat{r}} }{2} }^T$, and $e_t$ is the $t^{th}$ column of the identity matrix. The update in Eq. \ref{EQ:GPAL:GFS:IGPR:UP} allows $K_{\PAREN{\breve{r}}}^{-1}$ to follow from the modified \emph{Woodbury} formulation \cite{SAIGAL} given by
\begin{equation}\label{EQ:GPAL:GFS:IGPR:INVUP}
\displaystyle
\begin{split}
K_{\PAREN{\breve{r}}}^{-1} & = \bar{K}^{-1} + d_u \bar{u}\bar{u}^T - d_v \bar{v}\bar{v}^T, \quad \bar{K}^{-1} = \BRAK{\begin{array}{cc}
K_{(r)}^{-1} & 0 \\
0 & 1
\end{array}},\\
\bar{u} & = \bar{K}^{-1}u, \quad d_u = \PAREN{1-<\bar{u}, u>}^{-1}, \\
\bar{v} & = \PAREN{\bar{K}^{-1} + d_u \bar{u}\bar{u}^T}v, \quad d_v = \PAREN{1+<\bar{v}, v>}^{-1},
\end{split}
\end{equation}
which requires only two rank-$1$ updates. For a fixed set of test inputs $X_*$, the updated posterior mean vector remains a matrix-vector product and the posterior variances are sums of diagonals given by 
\begin{equation}\label{EQ:GPAL:GFS:IGPR:POST}
\displaystyle
\begin{split}
\bar{f}_{*\breve{r}} & =  K_{*\breve{r}}K^{-1}_{(\breve{r})}Y_{\breve{r}}, \quad s_u  = K_{*\breve{r}} \bar{u}, \quad s_v = K_{*\breve{r}} \bar{v},  \\
\diag{\bar{\Sigma}_{*\breve{r}}} & = \diag{\bar{\Sigma}_{*r} + k_{*\hat{r}} k_{* \hat{r}}^T + d_u s_u s_u^T- d_v s_v s_v^T},
\end{split}
\end{equation}
where matrix $K_{*\breve{r}}=K(X_*,X_{\breve{r}})$.  The updated log-determinant is given by  $\log{\ABS{K_{\PAREN{\breve{r}}}}} = \log{\ABS{\bar{K}}} - \log{d_u d_v}$.  The total computational costs of updating the GP prior and posterior distributions at iteration $t$ are $\BIGO{t^2}$ and $\BIGO{N_* t}$ operations respectively.

\subsection{GP $L^2$ Risk Function Criterions} 
\label{SEC:GPAL:GFS:RISK}

We show how several risk functions can be derived from the $L^2$ distance between any two GP posterior mean functions evaluated at a possibly infinite sized set of test inputs $X_*$. Given two GPs $f_a$, $f_b$ defined over the subsets of inputs $X_a, X_b$ for indices $a$ and $b$, the $L^2$ distance between their two GP posterior mean functions ($\bar{f}_a = K_{*a} \hat{K}^{-1}_{a} Y_{a}$ and $ \bar{f}_b = K_{*b} \hat{K}^{-1}_{b} Y_{b}$) is analytic under certain GP prior assumptions. For prior mean $m(x)=0$ and the product of identical Mat\'{e}rn class covariance functions in Eq. \ref{EQ:GPAL:GPR:COV}, the errors evaluated at $X_*$ are given by
\begin{equation} \label{EQ:GPAL:GFS:RISK:LTWO}
\displaystyle
\begin{split}
\LTWOF{\bar{f}_a}{\bar{f}_b}{X_*}& = \sum_{x_* \in X_*} (\bar{f}_a - \bar{f}_b)^2 \\
& =  z_{a}^T Q_{aa}  z_{a} 
- 2z_{a}^T Q_{ab}  z_{b} 
+ z_{b}^T Q_{bb}  z_{b},
\end{split}
\end{equation}
where vectors $z_{a}  = \hat{K}^{-1}_{a} Y_{a}  \in \field{R}^{N_a} $, $z_{b}  = \hat{K}^{-1}_{b} Y_{b}  \in \field{R}^{N_b}$ are computed over training data. Updating the risk function evaluations between successive iterations $t$ is efficient as updating $\bar{f}_{a}$, $\bar{f}_{b}$ need only rank-$1$ updates via Eq. \ref{EQ:GPAL:GFS:IGPR:INVUP}. The associated matrices $Q_{ab}$, $Q_{aa}$, $Q_{bb}$ in Eq. \ref{EQ:GPAL:GFS:RISK:LTWO} are sub-matrices of $Q_{XX}$ and can be pre-computed in $\BIGO{N^2}$ operations. Computing $Q_{XX}$ depends on the following cases.

\textbf{Finite Case:}
If $X_*$ is finite, then matrices $Q_{aa}  =  \sum_{x_* \in X_*} K_{a *} K_{* a }   \in \field{R}^{N_{a} \times N_{a}} $, $Q_{ab} =  \sum_{x_* \in X_*} K_{a*} K_{* b } \in \field{R}^{N_{a} \times N_{b}} $, and $Q_{bb} =  \sum_{x_* \in X_*} K_{b *} K_{* b } \in \field{R}^{N_{b} \times N_{b}}$ are the summation of outer-products whose $i,j^{th}$ entries are products of Mat\'{e}rn class covariance functions in Eq. \ref{EQ:GPAL:GPR:COV}. 

\textbf{Infinite Case:}
If $X_* = (-\infty, \infty)$ is the full (unbounded) input domain, then matrices $Q_{aa}  =  \int^{\infty}_{-\infty} K_{a *} K_{* a } dx_*  \in \field{R}^{N_{a} \times N_{a}} $, $Q_{ab} =  \int^{\infty}_{-\infty} K_{a*} K_{* b } dx_*  \in \field{R}^{N_{a} \times N_{b}} $, and $Q_{bb} =  \int^{\infty}_{-\infty} K_{b *} K_{* b } dx_*  \in \field{R}^{N_{b} \times N_{b}} $ contain improper integral entries. For a valid distance measure, the posterior mean functions converge to identical zero-mean priors at the limits $x_{*k} \rightarrow \pm \infty$ and the improper integrals of the form $Q_{a_i b_j} = \prod_{k=1}^{D} F_{\nu i j k}$ given by
\begin{equation}\label{EQ:GPAL:GFS:RISK:F}
\displaystyle
\begin{split}
F_{\nu i j k} = \int^{\infty}_{-\infty} K_{\nu}(|x_{a_ik}-x_{*k}|, \ell_{a k}) K_{\nu}(|x_{b_j k}-x_{*k}|, \ell_{b k}) d x_{*k},
\end{split}
\end{equation}
are shown to be finite (see Appendix Eq. \ref{EQ:GPAL:APP:F}).  Several combinations of the $L^2$ distance are summarized as follows.

\textbf{Prediction Error $\LTWOF{\bar{f}_{\PAREN{\breve{r}}}}{y}{X}$:}
The prediction error is taken between the GP posterior means $\bar{f}_{\PAREN{\breve{r}}}$ at test inputs $X_* = X$ and the known sample pairs $(X, Y)$. 

\textbf{Generalized Error $\LTWOF{\bar{f}_{\PAREN{\breve{r}}}}{\bar{f}_{\PAREN{X}}}{X_*}$:} 
The generalized error is taken between two GP posterior mean functions $\bar{f}_{\PAREN{a}}$ and $\bar{f}_{\PAREN{b}}$ evaluated at any finite $X_*$ (may be out-of-sample from $X$). For GFS, the two GPs are specified by subset-selected $a=\PAREN{\breve{r}}$ and the full set of inputs $b = \PAREN{X}$.

\textbf{Normalized Error $\LTWOF{\frac{\bar{f}_{\PAREN{\breve{r}}}}{\NORM{\bar{f}_{\PAREN{\breve{r}}}}}}{\frac{\bar{f}_{\PAREN{X}}}{\NORM{\bar{f}_{\PAREN{X}}} } }{(-\infty, \infty)}$:}
The normalized error or ''frequentist`` risk is taken between two normalized GP posterior mean functions ($\frac{\bar{f}_{\PAREN{a}}}{\NORM{\bar{f}_{\PAREN{a}}}}$ and $\frac{\bar{f}_{\PAREN{b}}}{\NORM{\bar{f}_{\PAREN{b}}}}$) evaluated at $X_* = (-\infty, \infty)$ given uniform probability distribution over $x_*$. The norm term $\NORM{f} = \sqrt{\int_{-\infty}^{\infty} f(x)^2 dx}$ is shown to be finite by setting either of the functions in Eq. \ref{EQ:GPAL:GFS:RISK:LTWO} to zero. The two GPs are specified on subset-selected $a=\PAREN{\breve{r}}$ and the full set of inputs $b = \PAREN{X}$.


\subsection{Experiments} 
\label{SEC:GPAL:GFS:EXP}

\begin{figure}[ht]
  \centering
\includegraphics[width=.49\textwidth]{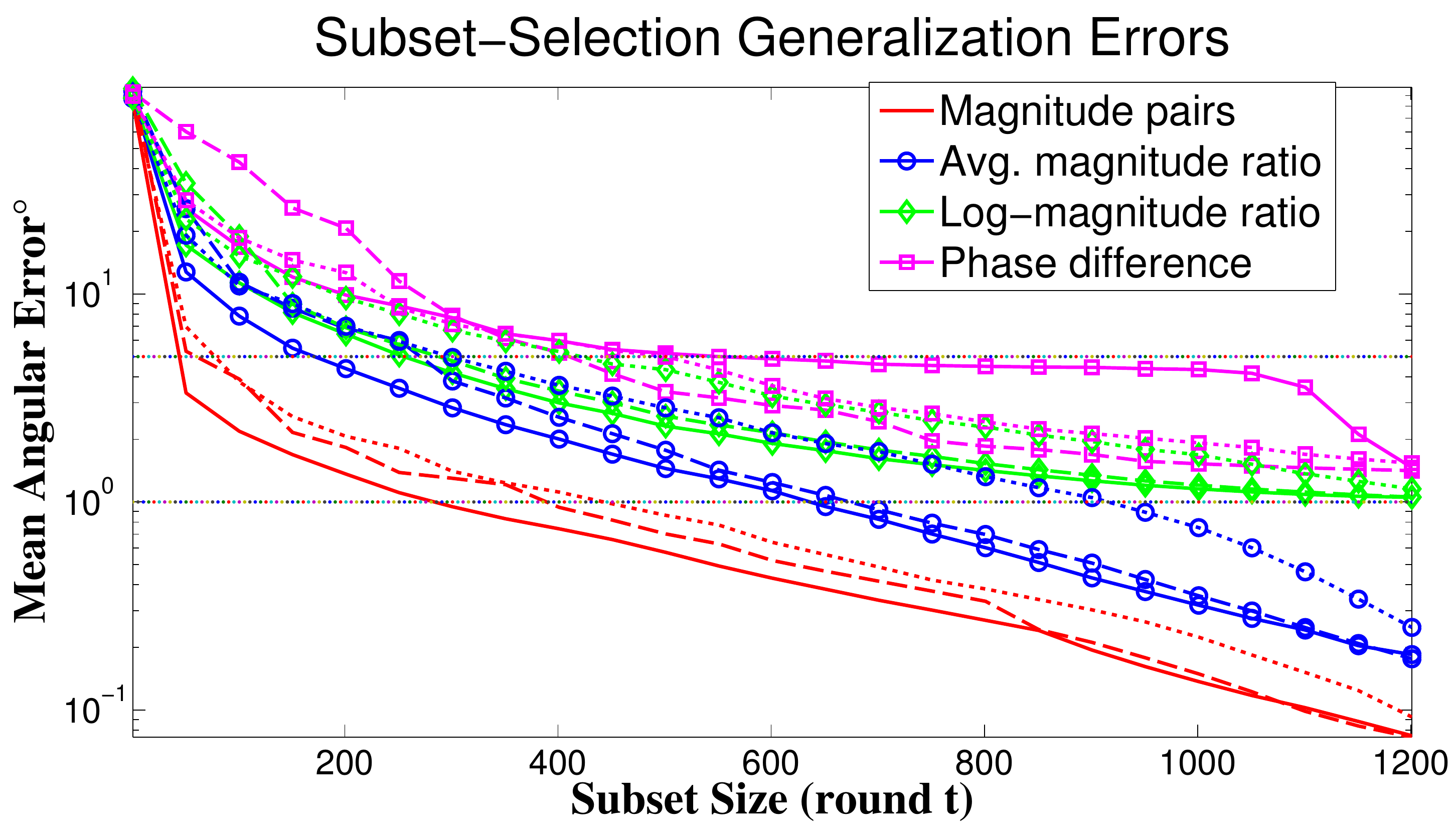}
\caption{Generalization errors are shown for GP-SSL models evidenced on randomized (dotted) and GFS [prediction error (solid), normalized error (dashed)] selected subsets of feature-direction pairs.}
\label{FIG:GPAL:GFS:EXP:MSS}
\end{figure}

GFS selects for increasing subset sizes until it contains the full dataset. At each iteration $t$, the incremental GP-SSL $K_{\infty}$ model infers directions (posterior means) along test inputs $X_*=X$. The mean angular separation error (Eq. \ref{EQ:GPAL:GPR:EXP:AS}) between the predicted and the reference measurement directions are computed and shown in Fig. \ref{FIG:GPAL:GFS:EXP:MSS}; intercepts with horizontal lines indicate subset sizes at $5^{\circ}$ and $1^\circ$ errors. The crossover points at the $5^\circ$ error line (localization accuracy) are achieved for MP and AMR features at a small fraction of the total input set (approximately $50$ and $150$ feature-direction pairs);  decreases in localization error after $50$ randomized samples becomes logarithmic with diminishing returns. Moreover, GFS selected models generalize better than that of randomized selection in all but the PD features; a visual (second column plots in Fig. \ref{FIG:GPAL:GPR:EXP:GEN}) shows that the former more accurately localizes directions further from the median plane.

\section{Active-Learner System}
\label{SEC:GPAL:ALM}

\begin{figure}[ht]
  \centering
\includegraphics[width=.45\textwidth]{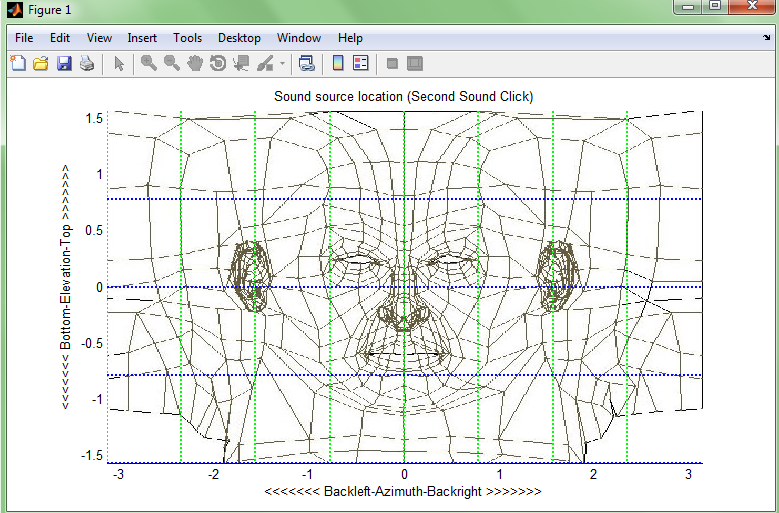}
\caption{GUI shows a mercator projection of spherical coordinate system onto $2$D panel. User clicks on panel to report a direction.
}\label{FIG:GPAL:ALM:GUI}
\end{figure}

The active-learning process for inferring HRTFs is as follows. The collection of $p$ number of target directions is specified as $\textbf{u} \in U \in \field{R}^{3 \times p}$. For rounds $t<T$, a query HRTF (MP) $\textbf{x}_t$ is chosen from the candidate set $X^C$ and appended to form input matrix $X \in \field{R}^{T \times D}$. The listener localizes  $\textbf{x}_t$, registers the direction $\textbf{v}_t$ over a GUI (see Fig. \ref{FIG:GPAL:ALM:GUI}), and appends the directions to form matrix $V \in \field{R}^{3 \times T} $.  The SSLEs w.r.t. $U$ are computed in $Y_{\textbf{u}t} = \textbf{SSLE}(\textbf{u}, \textbf{v}_t)$   s.t. $Y =  -U^T V \in \field{R}^{p \times T}$. Last, the updated feature-direction pairs $(X,Y)$ are added into the GP-SSLE models via incremental GPs (section \ref{SEC:GPAL:GFS:IGPR}). The system components are organized below.

\subsection{Conditional Mixture of Gaussians Models}
\label{SEC:GPAL:ALM:MOG}

While it is possible to specify an entire HRTF database as the candidate set, it is reasonable to assume that most samples would not be localized near a target direction $\textbf{u}$; overt features arising from the reflections off the anthropometry may be a physical impossibility along all measurement directions. Conversely, choosing only HRTFs with measurement directions equivalent to $\textbf{u}$ restricts the sample size to the number of subjects in the database. To address both issues, we model both the HRTFs and their corresponding measurement directions using a conditional mixture of Gaussians model (MoG) trained from the CIPIC database (see section \ref{SEC:GPAL:ALM:MOG}). This allows for $X^C$ to be drawn from a distribution of HRTFs conditioned at any direction $\textbf{u}$.

The MoG models the joint distribution between input variables as if the samples are drawn from a latent set of normal distributions. The input variables consist of measurement directions $\textbf{u}$ and leading principal components (PCs)\footnote{PCs are computed from same-subject, mean-centered, log-magnitude pairs (concatenated left and right ear HRTFs).} $\textbf{w}$ associated with HRTFs along $\textbf{u}$. The joint distribution is modeled by a weighted sum of $M$ normal distributions with mean and covariances given by
\begin{equation}\label{EQ:GPAL:ALM:MOG:GMM}
\displaystyle
\begin{split}
\textbf{z} &  = \BRAK{\begin{array}{c} \textbf{w} \\ \textbf{u}\end{array}}, \quad 
\mu = \BRAK{\begin{array}{c} \mu_\textbf{w} \\ \mu_\textbf{u}\end{array}}, \quad 
\Sigma = \BRAK{\begin{array}{cc}
\Sigma_{\textbf{w}} & \Sigma_{\textbf{w}\textbf{u}} \\
\Sigma_{\textbf{u}\textbf{w}} & \Sigma_{\textbf{u}}
\end{array}}, \\
P(\textbf{z}) & = \sum_{i=1}^M \pi_i \mathcal{N}\PAREN{\textbf{z}|\mu^{\CBRAK{i}}, \Sigma^{\CBRAK{i}}}, \quad 
 \sum_{i=1}^M \pi_i = 1, 
\end{split}
\end{equation}
where parameters $\mu, \pi, \Sigma$ are trained via the well-known expectation-maximization algorithm. The PCs $\textbf{w}$ conditioned on  $\textbf{u}$ is also a MoG given by
\begin{equation}\label{EQ:GPAL:ALM:MOG:CGMM}
\displaystyle
\begin{split}
P(\textbf{w} | \textbf{u}) & = \sum_{i=1}^M \frac{\pi_i \mathcal{N}\PAREN{\textbf{u} | \mu^{\CBRAK{i}}_\textbf{u}, \Sigma^{\CBRAK{i}}_\textbf{u} } \mathcal{N}\PAREN{\textbf{w}| \mu^{\CBRAK{i}}_{\textbf{w}|\textbf{u}}, \Sigma^{\CBRAK{i}}_{\textbf{w}|\textbf{u}}} }{\sum_{j=1}^M \mathcal{N}\PAREN{\textbf{u} | \mu^{\CBRAK{j}}_\textbf{u}, \Sigma^{\CBRAK{j}}_\textbf{u} }} ,
\end{split}
\end{equation}
where the conditional mean and covariance for the $i^{th}$ mixture are $\mu^{\CBRAK{i}}_{\textbf{w}|\textbf{u}}  = \mu^{\CBRAK{i}}_\textbf{w} + \Sigma^{\CBRAK{i}}_{\textbf{w}\textbf{u}} \Sigma^{\CBRAK{i}^{-1}}_\textbf{u} (\textbf{u} - \mu^{\CBRAK{i}}_\textbf{u})$ and $\Sigma^{\CBRAK{i}}_{\textbf{w}|\textbf{u}}  =  \Sigma^{\CBRAK{i}}_\textbf{w} - \Sigma^{\CBRAK{i}}_{\textbf{w}\textbf{u}} \Sigma^{\CBRAK{i}^{-1}}_\textbf{u} \Sigma^{\CBRAK{i}^T}_{\textbf{w}\textbf{u}}$ respectively. 
The candidate set $X^C$ is given by PCs randomly sampled from the conditional MoG\footnote{Leading $16$ PCs are sampled (via Gibbs sampling) from one of $M=64$ multivariate normal distribution (randomly selected by weight).} in Eq. \ref{EQ:GPAL:ALM:MOG:CGMM} and decoded into HRTFs to form the candidate set. The non-individualized (directional-averaged) HRTFs are approximated by the sum of the weighted conditional mixture means.

\subsection{GPs for Modeling SSLE}
\label{SEC:GPAL:ALM:GPSSLE}

GP-SSLE models ($f_{1:p}(X) = \CBRAK{f_1(X), \hdots f_p(X)}$) are specified by a common set of input MP features $X$ and output SSLEs $Y$ for each of the $p$ number target directions in $U$. Accurate modeling of the SSLE depends on the choice of GP prior mean and covariance functions. A zero mean prior is reasonable as reported directions $\textbf{v}$ in the absence of localization should average to the zero vector. Choosing the GP covariance function is more difficult as the hyperparameters cannot be optimized in the absence of observations; inaccurate priors would result in poor generalizations error.

Fortunately, GP-SSLE models can be related to GP-SSL models when $U$ is the infinite set of target directions uniformly sampled over a unit sphere. Substituting the SSLE labels $Y = -U^T V$ into Eq. \ref{EQ:GPAL:GPR:COV:LOGLIKEGRAD}, the GP-SSLE LMH is now given by
\begin{equation}  
\label{EQ:GPAL:ALM:SPHERE:LOGLIKEGRAD}
\displaystyle 
\begin{split}
L & = - \frac{1}{2} \left ( |U|\log|\hat{K}| + \trace{Q U U^T} + t|U|
\log (2\pi) \right ),
\end{split}
\end{equation}
where  matrix $Q = V\hat{K}^{-1}V^T$. As $p \rightarrow \infty$, the sample covariance of $U$ approaches a constant variance $UU^T = \frac{1}{3}I$ due to symmetry. The LMH in Eq. \ref{EQ:GPAL:ALM:SPHERE:LOGLIKEGRAD} reduces to
\begin{equation}  \label{EQ:GPAL:ALM:SPHERE:LOGLIKEGRADSPHERE}
\displaystyle 
\begin{split}
L_S & = - \frac{|U|}{2} \left ( \log|\hat{K}| + \frac{\trace{Q}}{3} + t\log (2\pi) \right ),
\end{split}
\end{equation}
which is equivalent to that of GP-SSL models for MP features $X$ and directions $V$.  

\begin{figure}[ht]
  \centering
\includegraphics[width=.49\textwidth]{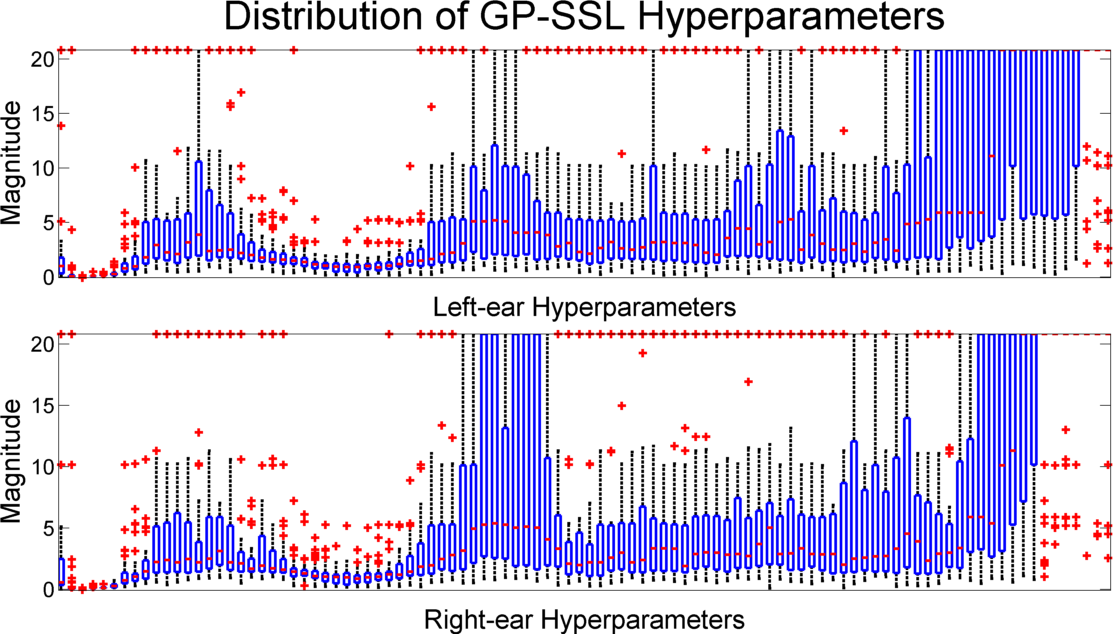}
\caption{Distribution (box-plot) of hyperparameter values are shown for GP-SSL models (x-axis $0-22.1$ kHz frequency range). Large valued hyperparameters $\ell_k$ indicate less sensitivity along the $k^{th}$ frequency.
}\label{FIG:GPAL:ALM:SPHERE:HYPER}
\end{figure}

The equivalence allows for the choice of the GP-SSLE model's covariance function to approximated by that of GP-SSL models trained over known feature-direction pairs (e.g. CIPIC subject data). While these subjects are not identical to the listener, the trained GP-SSL models all share similar covariance functions as their hyperparameters are well-distributed (see Fig. \ref{FIG:GPAL:ALM:SPHERE:HYPER}); high frequency bands above $17$ kHz tend to be negligible while lower frequency sub-bands between $0-3$ and $4-7.5$ kHz are relevant. 

%

\subsection{Query-Selection}
\label{SEC:GPAL:ALM:GPAL}

We present GP based query-selection as a modification of a known algorithm \cite{OSBORNE} which is derived as follows. Consider the observed minimum SSLE for any $\textbf{u}$ at round $t$ given by 
\begin{equation}
\label{EQ:GPAL:ALM:GPAL:GLOBAL}
\displaystyle
\begin{split}
\eta_{\textbf{u}t} = \min(Y_{\textbf{u}1}, \hdots, Y_{\textbf{u}t}).
\end{split}
\end{equation}
Realizations of SSLEs ($\gamma = f(\textbf{x}_* | X, Y)$) by the GP-SSLE posterior distribution  (Eq. \ref{EQ:GPAL:GPR:PREDICTION}) at a candidate input $\textbf{x}_* \in X^C$ will be normally distributed whose mean and variances represent the expected SSLE and uncertainty respectively. Thus, improvements (lowering) upon the global minimum $\eta_{\textbf{u}t}$ is given by the loss-function $\lambda_{\textbf{u}t}(\gamma) = \min(\gamma, \eta_{\textbf{u}t})$ whose expectation can be computed via marginalizing over the $\gamma$.

The expected loss-function is analytic for any single $\textbf{u}$ and so the weighted expected loss function (specified over each $\textbf{u} \in U$ with independent GP-SSLE models) is given by
\begin{equation}
\label{EQ:GPAL:ALM:GPAL:EL}
\displaystyle
\begin{split}
\wedge(x_*) & =  \sum_{\textbf{u} \in U} \rho_{\textbf{u}} \int_{-\infty}^{\infty}{\lambda_{\textbf{u}t}(\gamma) \mathcal{N}(\gamma|\bar{\mu}_\textbf{u}, \bar{C}_\textbf{u} )} d\gamma = \sum_{\textbf{u} \in U} \rho_{\textbf{u}} W_\textbf{u}, \\
W_\textbf{u} & = \eta_{\textbf{u}t} + (\bar{\mu}_\textbf{u}-\eta_{\textbf{u}t}) \psi(\eta_{\textbf{u}t}| \bar{\mu}_\textbf{u}, \bar{C}_\textbf{u}) - \bar{C}_\textbf{u} \mathcal{N}(\eta_{\textbf{u}t}| \bar{\mu}_\textbf{u}, \bar{C}_\textbf{u}),
\end{split}
\end{equation}
where weights $\rho_{\textbf{u}} = 1/p$ can be set to a constant, GP-SSLE posterior mean and covariance functions at $\textbf{x}_*$ evidenced with $(X_{1:t, :}, Y_{\textbf{u}, 1:t})$ are denoted by $\bar{\mu}_\textbf{u}$ and $\bar{C}_\textbf{u}$, and the cumulative normal distribution of $\bar{C}_\textbf{u}$ is denoted by $\psi$.  The query HRTF is chosen as the lowest scoring candidate or minimizer $\textrm{argmin}_{x_* \in X^C} \wedge(x_*)$ of the criterion Eq. \ref{EQ:GPAL:ALM:GPAL:EL} which balances local improvement through the posterior mean term $(\bar{\mu}_i-\eta_t)$ with exploring uncertain predictions through the posterior variance term $\bar{C}_\textbf{u}$. The property is useful for proving the rate of convergence \cite{SRINIVAS} to the true solution in Eq. \ref{EQ:GPAL:INTRO:SSLE}.

\subsection{Experiments}
\label{SEC:GPAL:ALM:EXP}

\textbf{GP-SSL active-learning trials:} One method for fast and repeatable empirical validation substitutes the human listener for GP-SSL models trained on CIPIC subject data. Localizations at $\textbf{x}_*$ can be reported as either the GP posterior mean directions, or by sampling from the GP posterior distribution.  This allows for large subsets of $X^C$ to be efficiently evaluated with little time costs. For coherence, we limit the query-selection criterion in Eq. \ref{EQ:GPAL:ALM:GPAL:EL} to single target directions $\textbf{u}$ belonging to the CIPIC HRTF measurement directions (queries made for past $\textbf{u}$ are discarded). GP-SSLE's covariance hyperparameters are set to that of the GP-SSL mean hyperparameters (averaged across $45$ subject models); hyperparameters can be retrained after each round but is not necessary for improving the localization error. The variance term is set to $\sigma=0.05$. 

\begin{figure}[ht!]
  \centering
\includegraphics[width=.45\textwidth]{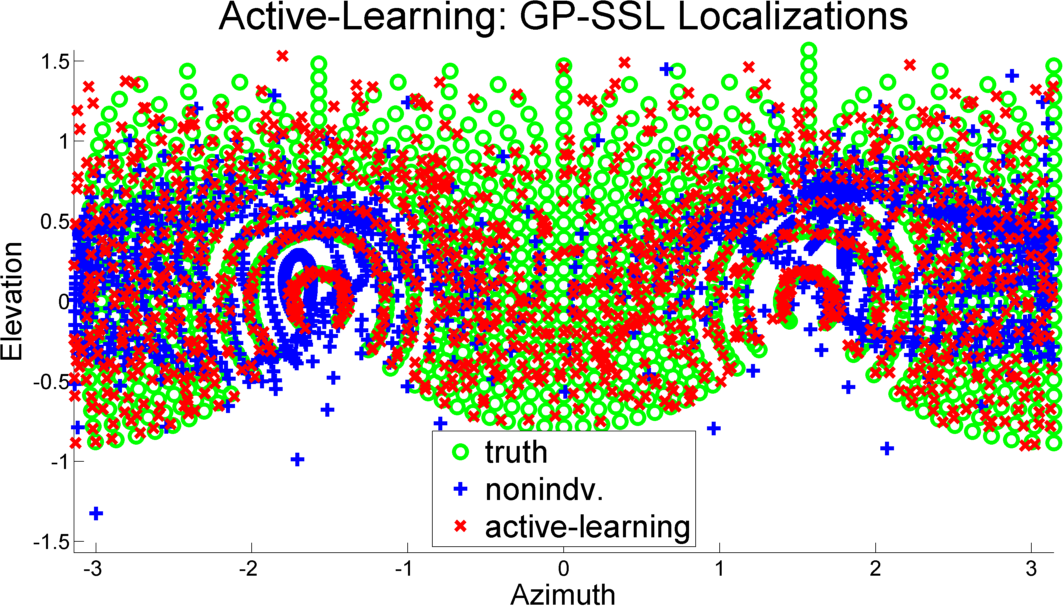}  
\caption{
Nearest localized directions after active-learning by the GP-SSL model (red) improve upon initial non-individualized HRTF localizations (blue).
}\label{FIG:GPAL:ALM:EXP:GPALM}
\end{figure}

\begin{figure}[ht!]
  \centering
\includegraphics[width=.49\textwidth]{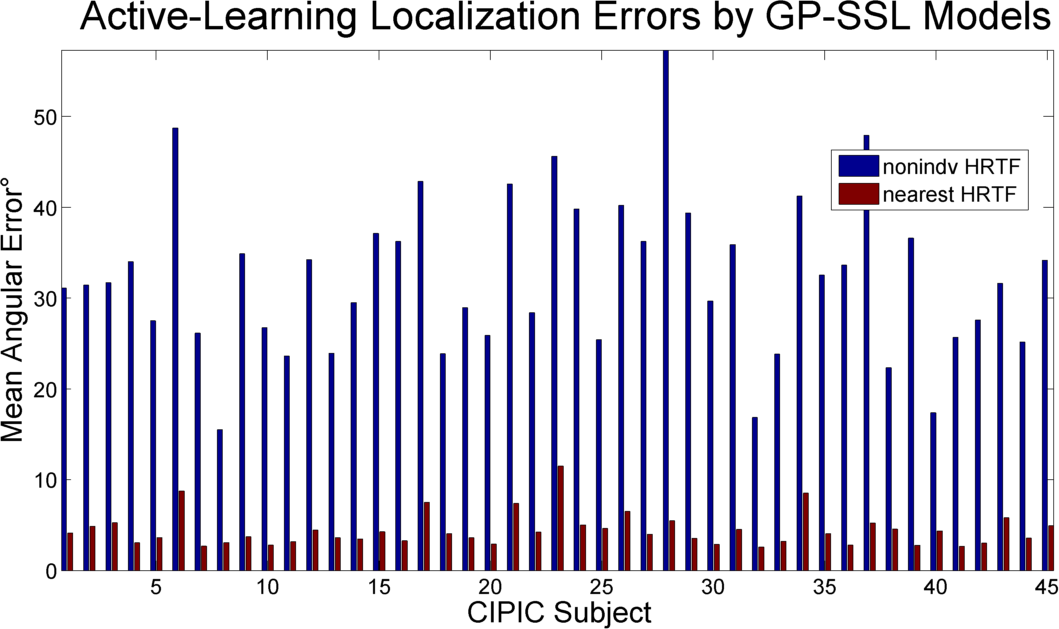} 
\caption{Mean angular errors are shown for the initial query (non-individualized HRTFs) and nearest HRTF queries.
}\label{FIG:GPAL:ALM:EXP:GPALMERR}
\end{figure}

In tests, the active-learner submits an initial non-individualized query HRTF for $\textbf{u}$ and then proceeds through $T=50$ rounds of query-selection from a candidate HRTF set of $20000$ samples drawn from a conditional MoG (Eq. \ref{EQ:GPAL:ALM:MOG:CGMM}). The nearest localized directions are shown to closer to their target directions than the non-individualized guesses (see Fig. \ref{FIG:GPAL:ALM:EXP:GPALM}). Non-individualized HRTFs are localized closer to the horizontal plane and towards the back of the head. Nearest localized directions accord with empirical studies of difficulties in front-back and up-down confusion with human subjects \cite{WENZEL2}. The experiment is repeated across all $45$ GP-SSL CIPIC subject models (see Fig. \ref{FIG:GPAL:ALM:EXP:GPALMERR}). The improvement can be expressed as the mean ratio between the angular separation errors of the initial and nearest localized directions. The mean improvement is $7.729$ across all CIPIC measurement directions, $9.139$ for median plane directions, and $8.252$ for horizontal plane directions. 

\textbf{Human active-learning trials:}
For a human listener, we develop a simple GUI in Matlab that consists of an azimuth-elevation plot that the subject clicks to report $\textbf{v}_t$. To introduce contrast in hearing, two test signals are alternatively played over headphones until the listener reports a direction. The first is a short burst of WGN independently generated for left and right ear channels. The second is the WGN convolved with the left and right min-phase HRTFs derived from the binaural MP features. The trials proceed as the listener localizes queries for $T=10$ rounds in each of the $14$ target directions ($7$ on the horizontal and median planes each).



For $5$ sample human listeners, the initial and nearest (minimum) localization errors for each of the target direction are shown in Table \ref{TAB:GPAL:ALM:EXP:PERDIR} and are compared to synthetic trials conducted with the $45$ GP-SSL CIPIC subject models. In both cases, the largest errors occur along the median plane direction $\theta= \CBRAK{-1.6, -0.69}$. The mean percentage improvements of the nearest localizations over that of the non-individualized HRTFs are  $49\%$ and $43\%$ for human and GP-SSL listeners respectively. GP-SSL localization errors are generally lower and more consistent across all direction than the human listeners; GP-SSL models can report a posterior mean direction whereas human listener exhibit variances in his/her localizations, even for identical test signals. It may be of interest in future work to both measure and model human localization variances via the GP-SSL's variance term $\sigma$ and by sampling localizations from the GP posterior distribution.

\begin{table}[ht]
\centering
\footnotesize
\caption{Active-learner: non-individualized and minimum horizontal $\phi$ and median $\theta$ plane localization errors (degrees)} \label{TAB:GPAL:ALM:EXP:PERDIR}
\begin{tabular}{|c|c|c|c|c|}
\hline
&\textbf{GP-SSL$_0$}&\textbf{GP-SSL$_{\min}$}&\textbf{Human$_0$}&\textbf{Human$_{\min}$}\\\hline
\textbf{$\phi: -2.4$}&$23.1 \pm 15.8$&$12.6 \pm 9.01$&$42.5 \pm 35.6$&$16.4 \pm 7.43$\\\hline
\textbf{$\phi: -1.6$}&$19.9 \pm 12.1$&$10.4 \pm 7.49$&$34 \pm 14.4$&$5.98 \pm 7.17$\\\hline
\textbf{$\phi: -0.79$}&$24.6 \pm 16.7$&$7.45 \pm 4.88$&$56.7 \pm 17.5$&$28.8 \pm 14$\\\hline
\textbf{$\phi: 0.79$}&$22 \pm 16.2$&$7.87 \pm 5.12$&$48.7 \pm 18$&$21.5 \pm 13.6$\\\hline
\textbf{$\phi: 1.6$}&$15.8 \pm 9.38$&$6.63 \pm 3.68$&$23.7 \pm 10.6$&$10.8 \pm 5.23$\\\hline
\textbf{$\phi: 2.4$}&$22.7 \pm 14.7$&$13.2 \pm 7.06$&$31.2 \pm 11.6$&$14.9 \pm 5.26$\\\hline
\textbf{$\theta: -1.6$}&$55.6 \pm 26$&$37.1 \pm 20.8$&$119 \pm 43.3$&$59.8 \pm 29.5$\\\hline
\textbf{$\theta: -0.79$}&$105 \pm 44.9$&$37.9 \pm 20.9$&$104 \pm 37.3$&$61.8 \pm 22.4$\\\hline
\textbf{$\theta: 0$}&$44.1 \pm 44$&$11.6 \pm 9.75$&$39.2 \pm 22.1$&$23.3 \pm 9.82$\\\hline
\textbf{$\theta: 0.79$}&$35.9 \pm 23.2$&$15.8 \pm 11.1$&$24.7 \pm 12.3$&$15.3 \pm 4.76$\\\hline
\textbf{$\theta: 1.6$}&$31.9 \pm 18.4$&$15.6 \pm 9.5$&$55 \pm 23.1$&$30.2 \pm 25.9$\\\hline
\textbf{$\theta: 2.4$}&$17.2 \pm 14.8$&$10.8 \pm 7.38$&$83.6 \pm 56$&$24.3 \pm 23.9$\\\hline
\textbf{$\theta: 3.1$}&$24.5 \pm 19.6$&$12.6 \pm 6.88$&$92.7 \pm 68.1$&$11.9 \pm 8.72$\\\hline
\textbf{$\theta: 3.9$}&$26.1 \pm 17.1$&$8 \pm 5.67$&$61.5 \pm 42.7$&$18.6 \pm 11.1$\\\hline
\end{tabular}
\end{table}

\section{Conclusions}

We developed a robust method for the SSL using sound-source invariant features derived from left and right ear HRTF measurements. Our GP-SSL models generalized NN based approaches and were shown to more accurate in both cases of randomized and subset-selected features; good spatialization accuracy ($5^{\circ}$) over the full sphere was possible using a fraction of the available features. For learning HRTFs in listening tests, we developed an active-learning method for query-selection using GP models. Both simulations with offline GP-SSL models and HRTFs recommended to real human listeners have shown large improvement in localization accuracy over non-individualized HRTFs.

\begin{appendices}

%

\section{Mat\'{e}rn Product Integrals}

Improper integrals in Eq. \ref{EQ:GPAL:GFS:RISK:F} have closed-formulations:
\begin{equation} \label{EQ:GPAL:APP:F}
\displaystyle
\begin{split}
F_{\frac{1}{2} i j k} & = \PAREN{\ell_{ak} e^{\frac{-\ABS{x_{a_ik}-x_{b_jk}}}{\ell_{ak}}}  - \ell_{bk} e^{\frac{-\ABS{x_{a_ik}-x_{b_jk}}}{\ell_{bk}}}} \frac{2\ell_{ak} \ell_{bk} }{\ell_{ak}^2 + \ell_{bk}^2 }, \\
F_{\frac{3}{2} i j k} & = \left ( \ell_{ak}^2(\ell_{ak}  - \beta \ell_{bk} - \alpha) e^{\frac{-\sqrt{3}\ABS{x_{a_ik}-x_{b_jk}}}{\ell_{ak}}} \right . \\  + & \left . 
\ell_{bk}^2(\ell_{bk}  + \beta \ell_{ak} - \alpha)  e^{\frac{-\sqrt{3}\ABS{x_{a_ik}-x_{b_jk}}}{\ell_{bk}}} \right )
\frac{4\ell_{ak}\ell_{bk}}{\sqrt{3}\PAREN{\ell_{ak}^2 - \ell_{bk}^2}^2}, \\
& \alpha = -\sqrt{3}\ABS{x_{a_ik} - x_{b_jk}}, \quad \beta = \frac{4\ell_{ak}  \ell_{bk}}{\ell_{ak}^2 - \ell_{bk}^2}, \\
F_{\infty i j k} & = e^{-\frac{\PAREN{x_{a_i k} - x_{b_j k}}^2}{2\PAREN{\ell_{a k}^2 + \ell_{b k}^2 } }} \frac{\ell_{a k} \ell_{b k} \sqrt{2 \pi}  }{ \sqrt{\ell_{a k}^2 + \ell_{b k}^2} }. 
\end{split}
\end{equation}

\end{appendices}

\bibliographystyle{IEEEtran}
\bibliography{masterbib}




\end{document}